\documentclass[pra,twocolumn,superscriptaddress,showpacs]{revtex4-1}
\usepackage{hyperref}
\usepackage{bbm}
\usepackage{amsmath, amssymb}
\usepackage{color}
\usepackage{graphicx,epsfig}
\usepackage{pifont}
\usepackage{subfigure}
\usepackage{amsmath, amssymb, graphics}

\newcommand{\kaptil}{\tilde{\kappa}}

\begin{document}

\title{Quantum correlations in the 1-D driven dissipative transverse field
  XY model.} 

\author{Chaitanya Joshi} \email{cj30@st-andrews.ac.uk}
\affiliation{Scottish Universities Physics Alliance, School of Physics
  and Astronomy, University of St Andrews, St Andrews KY16 9SS, United
  Kingdom}
\author{Felix Nissen}
\affiliation{London Centre for Nanotechnology, University College London, 17-19 Gordon St, London WC1H 0AH, United Kingdom}

\author{Jonathan Keeling} 
\affiliation{Scottish Universities Physics Alliance, School of Physics
  and Astronomy, University of St Andrews, St Andrews KY16 9SS, United
  Kingdom}

%\date{\today}
\begin{abstract}
  We study the non-equilibrium steady state (NESS) of a driven
  dissipative one-dimensional system near a critical point, and
  explore how the quantum correlations compare to the known critical
  behavior in the ground state.  The model we study corresponds to a
  cavity array driven parametrically at a two photon resonance,
  equivalent in a rotating frame to a transverse field anisotropic XY
  model [C. E. Bardyn and A. Imamo\u{g}lu, Phys. Rev. Lett {\bf 109}
  253606 (2012)].  Depending on the sign of transverse field, the
  steady state of the open system can be either related to the ground
  state or to the maximum energy state.  In both cases, many
  properties of the entanglement are similar to the ground state,
  although no critical behavior occurs. As one varies from the Ising
  limit to the isotropic XY limit, entanglement range grows.  The
  isotropic limit of the NESS is however singular, with simultaneously
  diverging range and vanishing magnitude of entanglement. This
  singular limiting behavior is quite distinct from the ground state
  behavior, it can however be understood analytically within
  spin-wave theory.
\end{abstract}

%\pacs{03.65.Ud, 03.75.Gg, 75.10.Pq}
\maketitle
\noindent
\section{Introduction}

% Ising model as paradigmatic QPT model
A central feature of critical behavior in any non-mean-field
phase-transition is the existence of a diverging correlation length
\cite{Sachdev2011,kadanoff2000statistical}.  Such divergences explain
why universal theories, controlled only by symmetries of the problem,
apply in the vicinity of a critical point.  They also lead to scaling
behavior~\cite{kadanoff2000statistical} of correlation functions.
More recently, it has been noted that measures of specifically
\emph{quantum} correlation,
e.g.\ entanglement~\cite{nielsen2010quantum}, also show scaling
behavior~\cite{Osterloh2002,Osborne2002,Vidal2003a,Amico2008}.
Entanglement is one of the characteristic traits of quantum
mechanics~\cite{nielsen2010quantum} and is of practical significance
as it captures quantum correlations which can be a resource for
quantum cryptography, quantum teleportation, and dense coding
\cite{Horodecki2009a}.  Despite the diverging correlation length at
critical points, entanglement generally has a finite
range~\cite{Osterloh2002,Osborne2002,Amico2008}, critical scaling is
instead seen in the magnitude of the entanglement.
 
% Open system and QPT, progress so far.
In a dissipative system, coupling to an external
environment~\cite{Breuer2002} leads to dephasing, and consequent
degradation of quantum correlations, ultimately reducing the system to
a classical description~\cite{Zurek2003,LoFranco2013}.  Nonetheless, in a
coherently driven dissipative system, i.e.\ pumped by an external
coherent drive, non-trivial steady states can be found
\cite{Dimer2007,Hartmann2010a,Baumann2010,Nagy2010,Diehl2010,Ferretti2010,Lee2011a,Marcos2012,Murch2012,Lee2012,Grujic2012,Torre2011,Torre2012,LeBoite2013a,Leib2013,Lee2013,Genway2013,Hu}.
In an extended interacting driven dissipative system, such as an array
of coupled nonlinear cavities as discussed below, this enables
non-local quantum correlations to exist in the non-equilibrium
steady state.  Such systems allow one to study quantum correlations
out of equilibrium, and to study whether dissipation has particular
significance for distinctively quantum correlations such as
entanglement.

% Aim of this paper.
The aim of this paper is to explore the range and scaling of quantum
correlations in the non-equilibrium-steady-state (NESS) near to a
critical point of the corresponding equilibrium system.
% Majorana modes and Ising model
A natural system in which to address such questions is an array of
coupled
cavities~\cite{Hartmann2006,Greentree2006,Angelakis2007a,Hartmann2008,Schmidt2013b}.
Such systems allow for tunable coupling and nonlinearity, and
inevitably have dissipation, as light escapes from the cavities.
Recently \citet{Bardyn2012} have shown that such systems can in
certain limits map to dissipative spin chain models, as explained
below.  Their proposed configuration allows tuning of both the
anisotropy of the spin-spin coupling, and of a transverse field.  We
study the non-equilibrium steady state, i.e.\ the long time behavior,
in the presence of dissipation.  Within this scenario, we determine
the dependence of quantum correlations on both of these parameters,
exploring the range from the transverse field Ising model to the
transverse field XY model.

The transverse field Ising model is a paradigmatic example of quantum
critical behavior~\cite{Sachdev2011}, and so the scaling of
entanglement in the equilibrium Ising model (or anisotropic XY model)
was one of the first examples
studied~\cite{Osterloh2002,Osborne2002,Vidal2003a}. As noted above,
while the magnitude of entanglement shows critical scaling, the range
over which non-zero entanglement exists does
not~\cite{Osterloh2002,Osborne2002}.  This finite range behavior
persists for all models in the Ising universality
class~\cite{Osborne2002,Maziero2010,StelmachoviPeter2004}.  Following
these early studies, there have been many subsequent explorations of
critical entanglement, including the spin-boson system~\cite{Hur2008}
which can be viewed as a phase transition of a dissipative quantum
system.  For a review, see Ref.~\cite{Amico2008}.

A major difficulty in understanding a many body quantum system is the
exponential growth of Hilbert space dimension with the system size.
One method to overcome this difficulty is to use a matrix product
state (MPS) approach~\cite{Vidal2003,Vidal2004}.  Such methods make
use of the fact that many physically relevant states have entanglement
which is either constant or grows at most polynomially with system
size~\cite{Zurek2003}; an MPS can efficiently represent such a state.
The MPS representation of a state is the concept underlying the
Density Matrix Renormalization Group
(DMRG)~\cite{White:DMRG,White:Algorithms} approach.  While the DMRG was
originally used as a method to determine ground states of interacting
systems, it was later extended to study dynamics
\cite{Cazalilla2002,White2004,Daley2004,Micheli2004,Clark2004,Cai2013}, by an
approach known as time evolving block decimation (TEBD).  All these
approaches ultimately rely on the fact that an efficient MPS
representation of the relevant states of the system exists, for a
discussion of this see e.g.  Ref.~\cite{Schollwock2011a}.  These
approaches have also been extended to open systems (mixed states), by
introducing matrix product operators
(MPO)~\cite{Zwolak2004,Orus2008,Prosen2009}.  This allows one to
efficiently time evolve the density matrix equations of motion for one
dimensional open systems, and thus find the non-equilibrium steady
state.

The remainder of this paper is organized as follows.
Section~\ref{sec:driv-diss-ising} reviews the basis of our
calculations.  In particular,
sections~\ref{sec:effect-hamilt},\ref{sec:dissipation} introduce the
effective Hamiltonian we study and its coupling to an external
environment; section~\ref{sec:meas-quant-corr} reviews the measures of
quantum correlations we calculate;
section~\ref{sec:matrix-product-state} outlines the MPO method we use
to find the steady state.  Section~\ref{sec:scal-quant-corr} then
presents the results of our numerical calculation.  After reviewing
the nature of the steady state in section~\ref{sec:nature-non-equil},
and comparing these results to the mean-field theory of our model
  in section ~\ref{sec:mft}, 
sections~\ref{sec:open-syst-corr},\ref{sec:open-syst-corr-1},\ref{sec:corr-vs-decay}
discuss the dependence of quantum correlations on each of the model
parameters in turn.  Finally, section~\ref{sec:asymptotic-delta-to}
discusses analytic calculations which can reproduce the behavior seen
for weak driving.  In section~\ref{sec:conclusions} we summarize our
findings.

\section{Driven-dissipative model and observables}
\label{sec:driv-diss-ising}

\subsection{Effective Hamiltonian}
\label{sec:effect-hamilt}

\begin{figure}[htpb]
  \centering
  \includegraphics[width=3.2in]{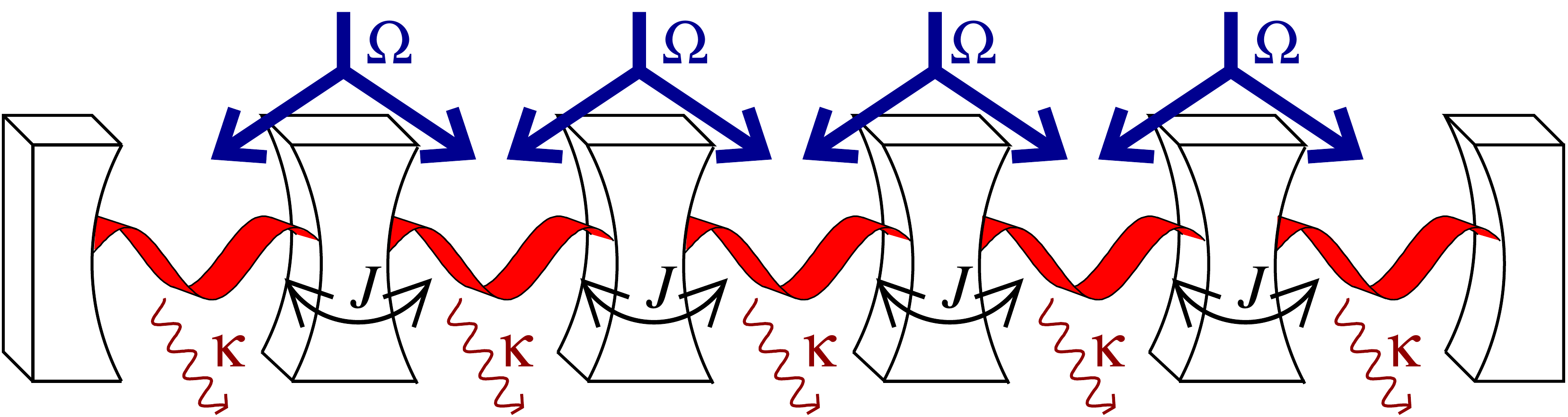}
  \caption{(Color online) Cartoon illustrating coupled cavity array with hopping $J$,
    two-cavity pumping $\Omega$ and loss rate $\kappa$.}
  \label{fig:cca-cartoon}
\end{figure}

We consider a coupled cavity array realization of the transverse field
anisotropic XY model, as introduced in Ref.~\cite{Bardyn2012}.  For
completeness, we briefly summarize the nature of such a model here.
As illustrated in Fig.~\ref{fig:cca-cartoon} the model consists of a
1-D array of optical cavities, supporting photon modes, described by
bosonic operators $c_j$ with hopping amplitude $J$ between the
cavities so that $H = \sum h_j - J \sum_{j} [ c^\dagger_j c^{}_{j+1} +
\text{H.c.} ]$.  The on-site Hamiltonian, $h_j= \omega_c c^\dagger_j
c^{}_j + U c^\dagger_j c^\dagger_j c^{}_j c^{}_j $, incorporates an
optical nonlinearity $U$. Physically this can be induced by coupling
each cavity to a saturable optical
absorber~\cite{Hartmann2006,Hartmann2008,LeBoite2013a}.

In addition to these elements, which would lead to a Bose-Hubbard
model~\cite{Fisher1989}, we include a two-photon driving term as
proposed in Ref.~\cite{Bardyn2012}.  Specifically, we consider a drive
$\Omega \cos(2 \omega_p t)$ near two-photon resonance, i.e.\ $\omega_p
\simeq \omega_c$, and we work in the limit of strong optical
nonlinearity.  In this limit, the problem simplifies, as one may
truncate each site to occupations $0$ or $1$.  Furthermore, this
implies that the two-photon pump is only resonant for the creation of
pairs of photons on adjacent cavities.  When restricted to the $0,1$
occupation subspace, one may replace each cavity mode with a spin
$1/2$, i.e.  replace bosonic operators by Pauli matrices $(c^{}_j,
c^\dagger_j) \to (\sigma^-_j, \sigma^+_j)$. Here $\sigma^{\pm}_j =
(\sigma^x_j \pm i \sigma^y_j)/2$ in terms of regular Pauli matrices.
In this notation, the Hamiltonian becomes:
\begin{multline}
  \label{eq:1}
  \hat{H}_0 = \sum_j \frac{\omega_c}{2} \sigma_j^z
  - J \sum_j \left(\sigma^+_j \sigma^-_{j+1} + \sigma^-_j \sigma^+_{j+1} \right)
  \\
  - \Omega 
  \sum_j \left(\sigma^+_j \sigma^+_{j+1} e^{- 2 i \omega_p t} 
    +
    \sigma^-_j \sigma^-_{j+1} e^{2 i \omega_p t}  \right).
\end{multline}
The explicit time dependence appearing here can be removed by a transformation
to a rotating frame. In such a frame the Hamiltonian is given by:
\begin{multline}\label{eqnham2}
  \hat{H} = -J  \sum_{j} \left[ g \sigma_{j}^{z} +
  (\sigma_{j}^{+}\sigma_{j+1}^{-}+\sigma_{j+1}^{+}\sigma_{j}^{-})
  \right.
  \\
  \left.
  + \Delta 
  (\sigma_{j}^{+}\sigma_{j+1}^{+}+\sigma_{j+1}^{-}\sigma_{j}^{-}) \right],
\end{multline}
where we have introduced dimensionless parameters $g = (\omega_p -
\omega_c)/2J$ and $\Delta=\Omega/J$.   This can also be written
in the canonical form of the Ising model~\cite{mattis2006theory}:
\begin{equation}
  \label{eq:2}
  \hat{H} = - J \sum_j \left[
    g \sigma_{j}^{z}
    +
    \frac{1+\Delta}{2} \sigma_{j}^{x}\sigma_{j+1}^{x}
    +
    \frac{1-\Delta}{2} \sigma_{j}^{y}\sigma_{j+1}^{y}\right].
\end{equation}
The parameter $\Delta$ describes the anisotropy of the interaction:
$\Delta=0$ corresponds to the isotropic $XY$ model, $\Delta=1$ to the
Ising model.  For $0 < |\Delta| \le 1 $ the Hamiltonian is in the
Ising universality class.  In the ground state, changing the
transverse field $g$ will induce a quantum phase
transition~\cite{Sachdev2011} at $|g|=1$, between a phase with
$\langle \sigma_x \rangle \neq 0$ for $|g|<1$, and a phase with
vanishing $\langle \sigma_x \rangle$ for $|g|>1$.

\subsection{Dissipation}
\label{sec:dissipation}

In addition to the terms described so far, each cavity is also assumed
to couple to a continuum of radiation modes describing irreversible
loss into the environment~\cite{Breuer2002}.  At optical cavity and
pump frequencies, one may eliminate such modes via the Born-Markov
approximation~\cite{scully97,Breuer2002}, producing the master equation:
\begin{equation}\label{equationfin}
  \frac{d}{d t}\rho =-i[\hat{H},\rho]  
  +\kappa \sum_{j} [2
  \sigma_{j}^{-}{\rho} \sigma_{j}^{+}-\sigma_{j}^{+}\sigma_{j}^{-}\rho-
  \rho\sigma_{j}^{+}\sigma_{j}^{-}].
\end{equation}
The dissipation described in Eq.~\eqref{equationfin} corresponds to
independent incoherent loss from each cavity.  In the spin
language, this corresponds to a process that flips the spin from up
to down.  Such dissipation corresponds to a zero temperature bath,
this is appropriate when considering optical frequency systems as the
characteristic energy scales are much larger than temperature.  In the
following we introduce the dimensionless $\kaptil=\kappa/J$, and
consider the steady state of the system as a function of the
parameters $(g, \Delta, \kaptil)$.  In the remainder of the manuscript
all energies are thus given in units of $J$.

It is important to note that the form of Eq.~(\ref{equationfin}) can
only follow from an originally time-dependent, i.e.\ pumped system.
For a time-independent system coupled to an external bath, a correct
treatment of the bath~\cite{Cresser1992a} must lead to a Master
equation which drives the system toward its thermal state.  Such
behavior is clearly required to be consistent with equilibrium
statistical mechanics.  The same is not however true of a time
dependent Hamiltonian --- in the rotating frame, coupling to the bath
need not satisfy detailed balance due to the ``extra'' time dependence
induced by the pump frequency \cite{Joshi2013}.  The crossover between these limits as
one varies $\omega_c, \omega_p$ while keeping $g$ constant is an
interesting question for future work.

\subsection{Measures of quantum correlations}
\label{sec:meas-quant-corr}

To quantify the \emph{quantum} correlations between different sites
requires some care, since a given pair of sites will in general be
entangled both with other sites and with the external environment.  As
such, it is important to use a measure of the quantum correlation
between a specific pair of sites.  The measure of pairwise
entanglement we will use will be negativity, $\mathcal{N}$ defined as:
\begin{equation}
  \label{eq:4}
\mathcal N= \text{max}(0,\sum_{i}^{4} |\lambda_{i}|-1),
\end{equation}
where $\lambda_{i}$ are the eigenvalues of the partially transposed
two-qubit density matrix $\rho_{AB}^{T_B}$~\cite{nielsen2010quantum},
where $T_B$ indicates transpose for system $B$.  According to the
Peres-Horodecki criterion~\cite{Peres1996,Horodecki1996} a (mixed)
state of a bipartite system is separable if the negativity is zero.
For any separable state, the density matrix would remain positive
under a partial transpose. In an entangled state a partial transpose
may produce a non-positive density matrix~\cite{Lupo2005}.  The
negativity as defined in Eq.~(\ref{eq:4}) is a measure of whether the
partial transpose produces negative eigenvalues. A non-zero value of negativity  serves both as a necessary and sufficient condition for the inseparability of a general two qubit state~\cite{Peres1996,Horodecki1996}.

For pure states entanglement is a sufficient measure of quantum
correlations and quantifies the ability of a state to act as a
resource for quantum computational speed-up \cite{Josza2003}.  For
mixed states separability (vanishing entanglement) does not in general
imply classicality~\cite{Ollivier2001,Henderson2001,Modi2012b} ---
computational speed-up for mixed state quantum computing can occur
without entanglement~\cite{Knill1998a}.  Such speed-up has been
attributed to the presence of non-zero quantum
discord~\cite{Ollivier2001,Henderson2001,Modi2012b,Knill1998a,Dakic2012a},
$\mathcal D$ defined~\cite{Modi2012b} as follows: Consider a bipartite
system AB in a state $\rho$, and a local measurement performed on
subsystem B with its result ignored. Such a  measurement will cause a
disturbance of subsystem $A$ for almost all states.  There is however
a class, $\Omega$, of states that is unchanged by such a measurement.
For such states $\chi\in\Omega$, known as ``classical-quantum''
states, one may write: $\chi=\sum_{i} p_{i} \rho_{A i} \otimes |i
\rangle_B {}_B\langle i |$, where $p_{i}$ is a probability distribution,
$\rho_{A i}$ is the marginal density matrix of A, and the states
$|i\rangle_B$ form an orthonormal set.  Geometric discord $\mathcal D$
is the distance between the state $\rho$ and the closest
classical-quantum state $\chi\in\Omega$.  Explicitly, for an arbitrary
mixed state $\rho$ of a $d \otimes d$ quantum system it is $\mathcal
D(\rho)=\frac{d}{d-1}\text{min}_{\chi \in \Omega} ||\rho-\chi||^{2}$,
where $||M||=\sqrt{\sum_{i} m_{i}^{2}}$ is the Hilbert-Schmidt norm of
the operator $M$ with eigenvalues $m_{i}$.

In the specific case of two-level systems (qubits), a closed form for
$\mathcal{D}$ exists~\cite{Modi2012b,Dakic2010} Writing the state of
two qubits as:
\begin{equation}
\rho=\frac{1}{4} \sum_{i,j=0}^{3} R_{ij}\sigma_{i} \otimes \sigma_{j},
 \qquad
 \mathbf{R} = 
 \left(
 \begin{array}{cc}
   1 & \mathbf{y}^T \\ \mathbf{x} & \mathbf{t} 
 \end{array}
 \right)
\end{equation}
where $\sigma^{0,1,2,3}_j = \left\{\mathbbm{1}_j, \sigma^x_j,
  \sigma^y_j, \sigma^z_j \right\}$, and $\mathbf{R}$ is given in block
structure above, one may then construct the $3 \times 3$
matrix $S=(1/4)(\mathbf{x}\mathbf{x}^T + \mathbf{t}\mathbf{t}^{T})$  from which
\begin{equation}
\mathcal D=2 \text{Tr}[S]-2 \lambda_\text{max}(S),
\end{equation}
where $\lambda_\text{max}(S)$ is the largest eigenvalue of the matrix
S.

\subsection{Matrix product state evolution}
\label{sec:matrix-product-state}

As noted above, to find the non-equilibrium steady state, we time
evolve Eq.~(\ref{equationfin}) using a matrix product operator
approach~\cite{Zwolak2004,Orus2008,Prosen2009}.  We here briefly
summarize the method used in our calculation. Further details of our
specific implementation can be found in Ref.~\cite{Nissen2013a}.

Our problem requires time-evolving the density matrix of a chain of
$N$ two-level systems.  This density matrix may be written in the
form:
\begin{equation}\label{supr}
  \rho=
  \sum_{\{ i_1, i_2, \ldots, i_N\}} c_{i_1, i_2, \ldots, i_N} \bigotimes_{j=1}^N \sigma_j^{i_j}
\end{equation}
with $\sigma^{0,1,2,3}_j$ as given earlier forming a basis for the
density matrix on each site.  The central point of the MPO approach is
to write the coefficients $c_{ i_1, i_2, \ldots, i_N}$ in terms of
matrices $\Gamma^{[j]i_j}$ and vectors $\lambda^{[j]}$ as follows:
\begin{multline*}
  c_{i_{1},i_{2},\ldots i_{N}}=
  \sum_{\{\alpha_{j}\}}\Gamma_{1,\alpha_{1}}^{[1]i_{1}}\lambda_{\alpha_{1}}^{[1]}\Gamma_{\alpha_{1},\alpha_{2}}^{[2]i_{2}}\ldots
  \\
  \Gamma_{\alpha_{j-2},\alpha_{j-1}}^{[j-1]i_{j-1}}\lambda_{\alpha_{j-1}}^{[j-1]}\Gamma_{\alpha_{j-1},\alpha_{j}}^{[j]i_{j}}\ldots\Gamma_{\alpha_{N-2},\alpha_{N-1}}^{[N-1]i_{N-1}}\lambda_{\alpha_{N-1}}^{[N-1]}\Gamma_{\alpha_{N-1},1}^{[N]i_{N}}.
\end{multline*}
The matrix $\Gamma^{[j]i_{j}}$, corresponding to basis component $i_j$
on site $j$, is a $\chi_{j-1} \times \chi_j$ matrix, and
$\lambda^{[j]}$ is a set of $\chi_j$ coefficients associated with the
bond between site $j$ and site $j+1$.  Here $\chi_j$ is the (integer)
bond dimension of the matrix associated with bond $j$.  If all
$\chi_j=1$ then one has entirely separable density matrix, i.e.\ $\rho
= \bigotimes \rho_j$, equivalent to a mean-field approximation.  If
$\chi_j$ are sufficiently large, any state can be written in the above
form --- the required size for our two-level-system density matrix is
$\chi_j = \text{min}(4^j, 4^{N-j})$.  To efficiently simulate such a
system we restrict $\chi_j< \chi_{\text{max}}$.  For a fixed
$\chi_{\text{max}}$, the size of computation scales linearly with
chain length.  Despite this, the representation is able to accurately
describe the full quantum dynamics in many problems.

To time-evolve the state, we follow the algorithm described in
Ref.~\cite{Vidal2004,Orus2008}.  The Master equation may be written in
a superoperator form, with the density matrix as a vector $\rho \to
|\rho\rangle$, so that $\partial_t |\rho\rangle = M |\rho \rangle$.
The superoperator $M$ can be decomposed as $M = \sum_j
M^{\text{pair}}_{j,j+1} $, with the one-site operations split
between the appropriate pair operators.  Evolution by a time step
$\delta t$ corresponds to propagating the coefficients
$\Gamma^{[j]i_{j}}, \lambda^{[j]}$ under the operator
$\exp(M^{\text{pair}}_{j,j+1} \delta t)$.  This is done by converting
the MPO representation for a given pair of sites into its explicit
form, evolving the pair, and then performing a singular value
decomposition (SVD)~\cite{nielsen2010quantum} to return the final form
to MPO representation.  The rank $\chi_j$ after such an update will
generally have increased, but can be restored to $\chi_j \le
\chi_{\text{max}}$ by keeping only the largest $\chi_{\text{max}}$
singular values in the SVD.

To extend from a single pair to many, the overall superoperator $M$
can be divided into parts on odd and even sites $j$ and, using the
Suzuki-Trotter expansion $e^{M \delta t} = e^{M^{\text{odd}} \delta
  t/2} e^{M^{\text{even}} \delta t} e^{M^{\text{odd}} \delta t/2} +
\mathcal{O}(\delta t^3)$.  Since $M^{\text{odd}}$ involves a sum of
pair operations which each mutually commute, all the updates in
$M^{\text{odd}}$ can be performed in parallel.  The same applies to
$M^{\text{even}}$. 

\begin{figure}[htpb]
  \centering
  \includegraphics[width=3.0in]{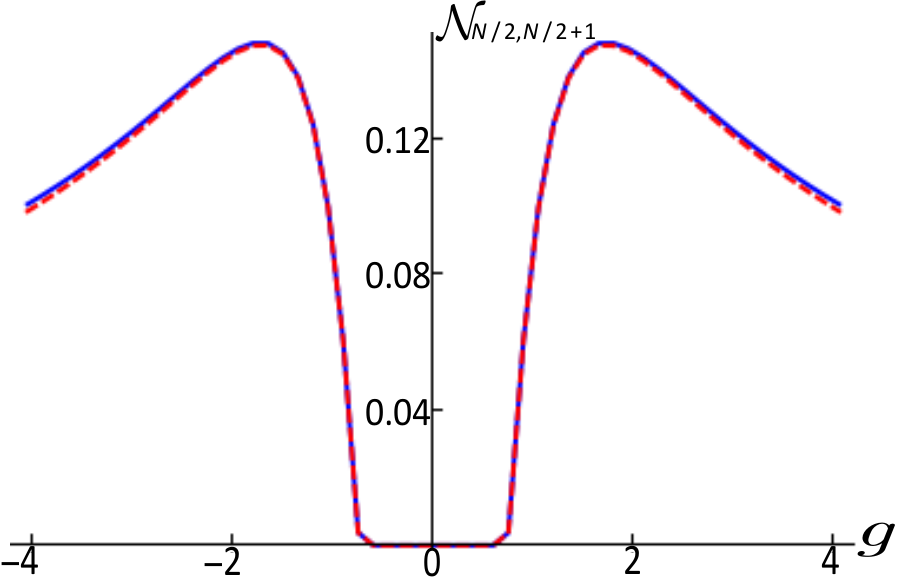}%{4site.pdf}
  \caption{(Color online) Negativity $\mathcal N$ vs transverse field
    $g$, comparing MPO numerical solution for $\chi_{\text{max}}=20$ (blue-solid)
    to exact diagonalization (red-dashed) for a four-site Ising
    model. Parameters (in units of $J$): $\Delta=1$, $\kaptil=0.5$.}
  \label{fig:compare-exact-mps}
\end{figure}

To demonstrate the accuracy of our implementation~\cite{Nissen2013a},
Fig.~\ref{fig:compare-exact-mps} shows a comparison between exact
diagonalization of the four-site Master equation and the open system
MPO code with $\chi_{\text{max}}=20$ showing close agreement. These
results, as all results in our paper, are calculated for a chain
with open boundary conditions. For
longer chains, comparison with exact solutions is not feasible so we
instead check for convergence of numerical results with matrix rank
$\chi_{\text{max}}$. Efficient simulation depends on whether
  convergence is achieved for sufficiently small values of the matrix
  rank $\chi_{\text{max}}$. If correlation lengths diverge, such as at
  critical points, strong long-range correlations exist. In such cases
  convergence would only occur at large $\chi_{\text{max}}$ and and
  evolution becomes computationally expensive.  In our system, we will
  see that the dissipation $\tilde{\kappa}$ suppresses such long range
  correlations; for small values of $\tilde{\kappa}$ the computational
  cost would increase, particularly near the equilibrium critical
  points $|g|=1$.  It is important also to note that in this paper we
  are only concerned with convergence of the steady-state properties.
  If one is also interested in the short time dynamics, the required
  matrix rank may be much larger~\cite{Hartmann2009}, due to transient
  correlations arising before dissipation has time to act. In
addition to convergence with matrix rank, we also find and check
that properties near the middle of the chain converge with increasing
chain length.

\section{Scaling of quantum correlations in Non-equilibrium steady
  states}
\label{sec:scal-quant-corr}

\subsection{Nature of the Non-Equilibrium Steady state}
\label{sec:nature-non-equil}

\begin{figure}[htpb]
  \centering
      \includegraphics[width=3.2in]{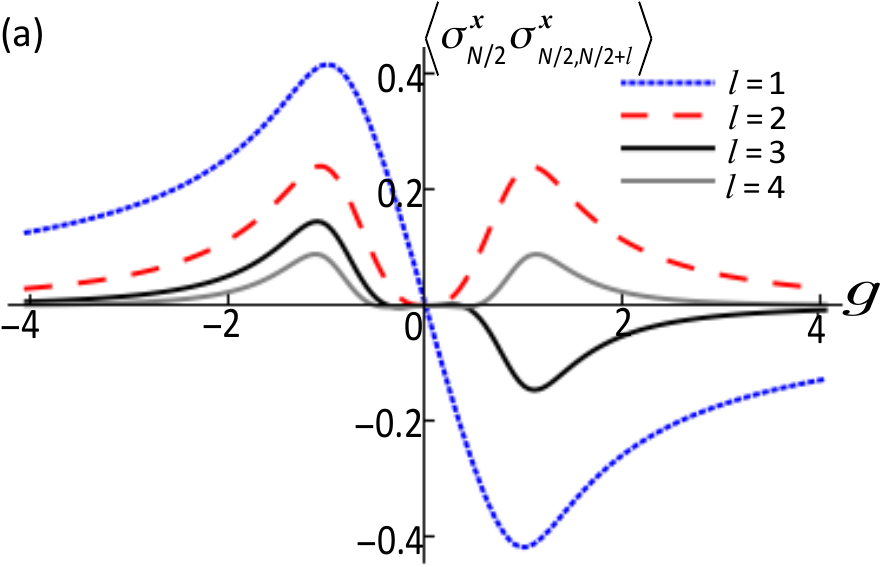}\\
        \includegraphics[width=3.2in]{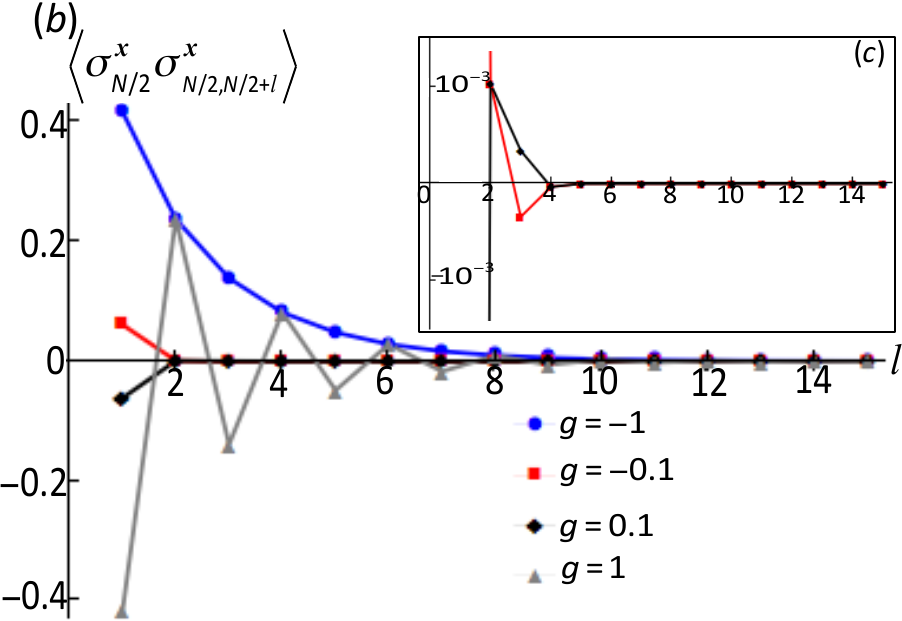}
        \caption{(Color online) Panel (a) showing spin-spin
          correlations $\langle \sigma^x_{j} \sigma^x_{j+l} \rangle$
          as a function of transverse field.  The different lines
          correspond to different separations.  Panel (b) showing
          decay of spin-spin correlations $\langle \sigma^x_{j}
          \sigma^x_{j+l} \rangle$ as a function of separation $l$
          between the spin sites. Both panels plotted for the Ising
          limit ($\Delta=1$). It is clearly seen that the NESS exhibit
          FM and AFM ordering for negative and positive values of
          transverse field ($g=\pm1$). Inset (panel (c)) shows short
          range incommensurate order for lower values of transverse
          field ($g=\pm 0.1$). The axes in the inset are same as in
          the main plot. Other parameters (in units of $J$):
          $\kaptil=0.5$ and MPO calculation performed for $N=40$ site
          chain, with $\chi_{\text{max}}=20$.  }%
   \label{fig:isingorder}
\end{figure}

Before discussing the quantum correlations in the non-equilibrium
steady state of Eq.~(\ref{equationfin}), we first discuss the nature
of the steady state itself.  The dissipation term on its own would
drive the system to a state with all spins pointing down. In the
  following we denote this state as the trivial empty state.  In
  general (unless $\Delta=0$), this trivial state is not an
eigenstate of the Hamiltonian so is not the steady state.  An
observable that gives a clear indication of the nature of the steady
state is the correlation function $\langle \sigma^x_{j} \sigma^x_{j+l}
\rangle$.  This is plotted in Fig.~\ref{fig:isingorder} for
$\Delta=1$, for sites near the center of the chain, hence avoiding
edge effects.

As is clear from Fig.~\ref{fig:isingorder}, in the NESS, the $x$
  components of spin show (short range) ferromagnetic order for
  transverse fields around $g\simeq -1$ and antiferromagnetic order for
  fields around $g\simeq 1$.  In comparison, in the ground state of
  the Ising model there are ferromagnetic correlations for $|g|<1$,
  regardless of the sign of $g$.  As will be proven below, there is a
  direct relation between the NESS for positive and negative $g$,
  corresponding to a $\pi$ rotation around the $z$ axis on every
  second lattice site.  This duality implies that if (short-range)
  ferromagnetic correlations are seen for a given $g$,
  anti-ferromagnetic correlations will exist for $g \to -g$.  As well
  as this formal duality, we will also discuss next a more intuitive
  picture for the different behaviors at positive and negative $g$, a
  picture substantiated by analytic results of mean-field analysis given in
  Sec.~\ref{sec:mft}.

For large negative $g$, the ground state is compatible with the
dissipation terms: both favor spins pointing in the $-z$ direction.
For weak decay ($\kappa \to 0$), steady states of the collective
dynamics generally correspond to stationary points of the
  closed-system dynamics.  Such stationary points will correspond to
  extrema of the energy. The ferromagnetic correlations seen for
$g<0$ clearly reflect the properties of the ground state, including a
peak in correlations near $g=-1$, where the ground state undergoes a
quantum phase transition.  In contrast, for large positive $g$ the
ground state is incompatible with the dissipation.  However, the
maximum energy state, which is also a stationary point of the
  dynamics is compatible with the dissipation.  The behavior of the
correlations seen in Fig.~\ref{fig:isingorder} suggests that for
$g<0$, the attractor of the dynamics is related to the ground state,
while for $g>0$ the attractor is instead related to the state of
maximum energy.  Similar behavior has been seen in the dynamics of the
Dicke model, where duality under change of sign of cavity-pump
detuning leads to an inverted normal state~\cite{Bhaseen:Noneqdicke}.

The proof of the duality under change of the sign of $g$ follows
  by considering transformations of the density matrix that relate its
  steady state for $g$ to that for $-g$.  We consider dividing the
chain into sublattices of odd and even sites.  The switch from ferro-
to antiferromagnetic order is equivalent to the statement that
correlations between sites on different (the same) sublattices are odd
(even) functions of the field $g$.  Two dualities are required to show
this.  Firstly, duality under $\hat{H} \to - \hat{H}$, $\rho \to
\rho^\ast$.  This follows from taking the complex (\emph{not}
Hermitian) conjugate of the equation of motion.  Since both $\hat{H}$
and all the loss terms are real, this complex conjugation means that
$\hat{H} \to - \hat{H}$ is equivalent to $\rho \to \rho^\ast$.  The
second duality concerns rotation around the $z$ axis on one
sublattice, $\rho \to \hat{R}_{\text{odd}} \rho \hat{R}_{\text{odd}}$
where $\hat{R}_{\text{odd}} =\prod_{j=1,3,5\ldots} \sigma^z_j$, this
has the effect of modifying Eq.~(\ref{eq:2}) by changing the sign of
the inter-site couplings; this is equivalent to the combination $H \to
-H, g \to -g$.  Combining this duality with complex conjugation, one
finds that interchanging $g \to -g$ alone is equivalent to $\rho \to
\hat{R}_{\text{odd}} \rho^\ast \hat{R}_{\text{odd}}$.  This
transformation swaps the sign of correlations between the two
sublattices as required.  The dualities involved make clear the r\^ole
of the inversion $H \to -H$ in relating the steady states for $g \to
-g$, corroborating the statement that the $g>0$ steady state is
related to the maximum energy state.

 As can be seen in Fig.~\ref{fig:isingorder}(c), for small values
  of $g$, correlations become small, and vanish as $g = 0$.  In the
  small $g$ regime these small short-range correlations are neither
  strictly ferromagnetic nor anti-ferromagnetic, but instead show an
  incommensurate ordering.  Such behavior occurs in a regime where
  the mean-field theory would predict the trivial state.
  (Note that in other models, mean-field theory can also predict
  incommensurate orderings~\cite{Lee2013}.)  As expected the spin-spin
  correlation functions always respects the sublattice dualities as
  discussed above.

  The appearance of the trivial state as an attractor at $g\to 0$,
  cannot be simply related to minimum or maximum energy states as in
  the earlier discussion.  Note also that the above dualities do not
  explain why the same-sublattice correlators, which are even
  functions of $g$, should vanish at $g=0$.  The state at $g=0$ can
  nonetheless be directly understood: at $g=0, \Delta=1$, the
effective magnetic field seen by any site points purely in the $x$
direction, and so the evolution combines precession around the $x$
axis with decay.  Consequently, the $x$ component of all spins
vanishes at this point. The correlators $\langle \sigma^y_{j}
\sigma^y_{j+l} \rangle$ (not shown) do not generally vanish at
$g=0$, but still show the odd--even symmetry discussed above.  For
$\Delta<1$ the $\langle \sigma^x_{j} \sigma^x_{j+l} \rangle$ do not
vanish at $g=0$ either; this is discussed further in
  Sec.~\ref{sec:open-syst-corr-1}.

 \subsection{Comparison with the mean-field theory}
 \label{sec:mft} To further understand the differences between
   the NESS and the ground state, we next discuss the mean-field
   prediction for the NESS. While mean-field theory incorrectly
   predicts long-range order in low dimensions, the nature of the
   order predicted is reflected by the full MPO numerics.  Within
   mean-field theory it is possible to give closed-form expressions
   for the phase boundary, and for the nature of the order anticipated for
   given values of $g, \Delta, \kappa$.  This provides further
   intuition for the differences between the NESS and the ground state.

   In  mean-field theory, the full density matrix is
   approximated as a product state (i.e. equivalent to restricting
   $\chi_{\text{max}}=1$ in an MPO simulation).  The equations of motion  
   then reduce to the following set of non-linear Bloch equations:
   \begin{align}
     \label{eq:5}
    \frac{d}{d t}\langle \hat{\sigma}_{j}^{x}\rangle
    &=\! -\kaptil \langle \hat{\sigma}_{j}^{x}\rangle
    \!+\!2g\langle \hat{\sigma}_{j}^{y}\rangle 
    \!-\!(1-\Delta)\langle \hat{\sigma}_{j}^{z}\rangle (\langle \hat{\sigma}_{j-1}^{y}\rangle+\langle \hat{\sigma}_{j+1}^{y}\rangle) 
    \nonumber\\
    \frac{d}{d t}\langle \hat{\sigma}_{j}^{y}\rangle
    &=\! -\kaptil \langle \hat{\sigma}_{j}^{y}\rangle
    \!-\!2g\langle \hat{\sigma}_{j}^{x}\rangle 
    \!+\!(1+\Delta)\langle \hat{\sigma}_{j}^{z}\rangle (\langle \hat{\sigma}_{j-1}^{x}\rangle+\langle \hat{\sigma}_{j+1}^{x}\rangle)
    \nonumber\\
    \frac{d}{d t}\langle \hat{\sigma}_{j}^{z}\rangle
    &=\! -2\kaptil (\langle \hat{\sigma}_{j}^{z}\rangle+1)-(1+\Delta)\langle \hat{\sigma}_{j}^{y}\rangle (\langle \hat{\sigma}_{j-1}^{x}\rangle+\langle \hat{\sigma}_{j+1}^{x}\rangle)
    \nonumber\\
    & \! +(1-\Delta)\langle \hat{\sigma}_{j}^{x}\rangle (\langle
    \hat{\sigma}_{j-1}^{y}\rangle+\langle
    \hat{\sigma}_{j+1}^{y}\rangle).
  \end{align}
  One may either directly time-evolve these equations to determine
  steady states, or attempt to analytically solve these equations in
  cases where the spatial dependence is relatively
  simple.   Below we  first present the analytical
  approach, and then discuss direct numerical evolution.

  It is clear from Eq.~(\ref{eq:5}) that the trivial state $\langle
  \hat{\sigma}_{j}^{x}\rangle=\langle \hat{\sigma}_{j}^{y}\rangle=0,
   \langle \hat{\sigma}_{j}^{z}\rangle=-1$ is always a fixed
  point, i.e. a steady state.  This trivial state does not break the
  $\mathbbm{Z}_{2}$ symmetry of  Eq.~(\ref{equationfin}) and so can also be
  referred to as a paramagnetic state\cite{Lee2013}.  While such a
  steady state always exists, this state need not always be stable to
  small fluctuations.  To test linear stability, one may linearize
  the equations of motion around the steady state, and consider
  plane-wave fluctuations of the form:
  \begin{displaymath}
    \left(\begin{array}{c}
        \langle \hat{\sigma}_{j}^{x}\rangle  \\
        \langle \hat{\sigma}_{j}^{y}\rangle  \\
        \langle \hat{\sigma}_{j}^{z}\rangle  
      \end{array}
    \right)
    = 
   - \left(
      \begin{array}{c}
        0 \\ 0 \\ 1
      \end{array}
    \right)
    +
   \sum_{k} \left(
      \begin{array}{c}
        x_k \\ y_k \\ z_k
      \end{array}
    \right)
    e^{-i \nu_k t - i j k}.
  \end{displaymath}
  The equations of motion then yield a secular equation for the
  frequencies $\nu_k$, with solutions
  \begin{equation}\label{kappaeqn}
    \nu_k = - i \tilde{\kappa}
    \pm
    2\sqrt{g^2+2 g \cos (k)+(1-\Delta^2) \cos^2(k)}.
  \end{equation}
  and $\nu_k=-2i\tilde{\kappa}$.
  The steady state is stable to such a plane wave fluctuation $k$ if
  $\Im[\nu_k] <0$, meaning such fluctuations exponentially decay.

It is clear that for $|\Delta|<1$, the trivial state is
stable at both $g \to 0$ and $g \to \infty$.  The trivial state can be
unstable at intermediate $g$.  For positive $g$, the most
unstable fluctuations have $\cos(k)=-1$, i.e AFM
fluctuations, whereas for negative $g$ FM fluctuations,
$\cos(k)=1$ are the most unstable.
In the Ising limit $\Delta=1$, one can write a simple expression for
the phase boundary, $\tilde{\kappa} = 2\sqrt{1-(g\pm1)^2}$, indicating
that for small enough $\tilde{\kappa}$ the normal state is unstable
near to $g=\pm1$.

 In addition to the trivial state one may consider the FM ansatz
  $\langle \hat{\sigma}_{j}^{x}\rangle=X, \langle
  \hat{\sigma}_{j}^{y}\rangle=Y, \langle
  \hat{\sigma}_{j}^{z}\rangle=Z$, or AFM ansatz $\langle
  \hat{\sigma}_{j}^{x}\rangle=(-1)^j X, \langle
  \hat{\sigma}_{j}^{y}\rangle=(-1)^{j}Y, \langle
  \hat{\sigma}_{j}^{z}\rangle=Z$, and then find $X,Y,Z$ by
  substituting these forms into Eq.~(\ref{eq:5}) and solving the
  resulting cubic equation.  One finds that for negative $g$, there is
  a non-trivial FM solution ($X,Y \neq 0$), which exists only when the
  trivial state is unstable.  (When the trivial state is stable, the
  cubic equation only has one real root corresponding to $X=Y=0,
  Z=-1$.) For $g>0$ the same statements apply to the AFM ansatz.
  Whenever these non-trivial solutions exist they can be shown to be
  stable.

\begin{figure}[htpb]
  \centering
            \includegraphics[width=3.2in]{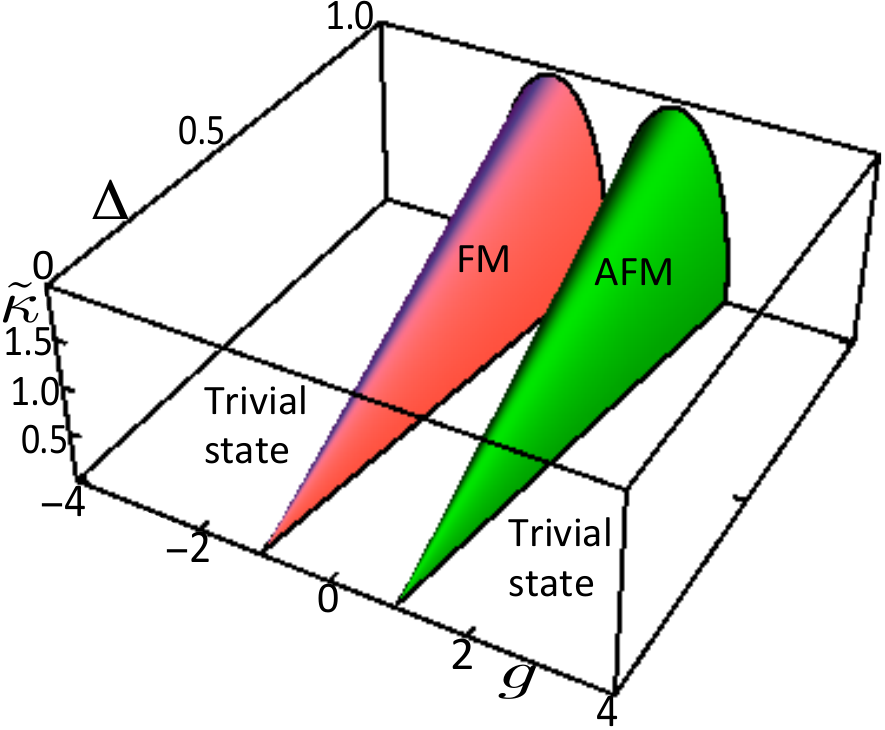}
      \caption{(Color online) Mean-field phase diagram for the
          non-equilibrium  steady state of Eq.~(\ref{equationfin}) 
 as a function of dimensionless parameters $g, \Delta, \kaptil$.}%
   \label{fig:meanfield}
 \end{figure}

 This analysis predicts a simple phase diagram, corroborated by direct
 numerical time evolution of Eq.~(\ref{eq:5}).  There are three
 phases, trivial, FM and AFM.  The
 boundaries between theses are given by the surfaces $\nu_\pi=0, \nu_0=0$ with
 $\nu_k$ from  Eq.~(\ref{kappaeqn}). This phase diagram is shown in
 Fig.~\ref{fig:meanfield} as a function of parameters $g,\Delta,
 \kaptil$.  It is clear  that for a fixed
 $\kaptil$ and with decreasing value of $\Delta$, the range of the
 transverse field strength $g$ over which the FM and AFM exist
 decreases.  As $\Delta \to 0$, for finite $\kappa$, the trivial
state  always occurs regardless of the value $g$.

\begin{figure}[htpb]
  \centering
      \includegraphics[width=3.2in]{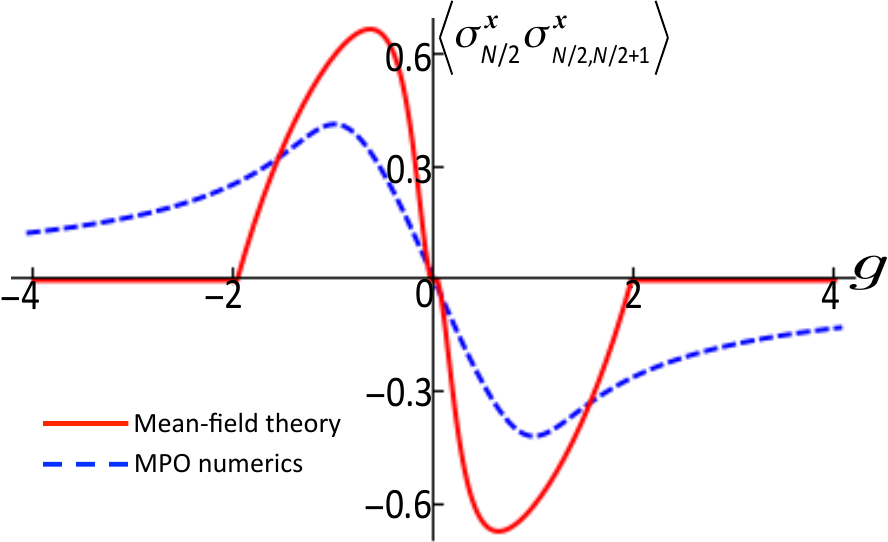}
      \caption{(Color online) Spin-spin correlations $\langle
        \sigma^x_{j} \sigma^x_{j+1} \rangle$ as a function of
        transverse field strength $g$. Parameters (in units of $J$):
        $\Delta=1, \kaptil=0.5$ and MPO calculation performed for
        $N=40$ site chain, with $\chi_{\text{max}}=20$.  }%
   \label{fig:comparemft}
\end{figure}

 To compare the predictions of mean-field theory and the full
 numerics, Fig.~\ref{fig:comparemft} compares their predictions for
 the correlation function $\langle \hat{\sigma}_{j}^{x}
 \hat{\sigma}_{j+1}^{x}\rangle$ as a function of transverse field
 strength $g$.  In the trivial state, MFT predicts this correlation to
 vanish, while in the ordered states it predicts $\pm X^2$, for the FM
 (AFM) states respectively. As can be seen, MFT does predict the kind of order that
 is seen, but predicts sharp phase boundaries that are not seen in the
 full numerics.

As noted above, direct time evolution of Eqs.~(\ref{eq:5}) corroborate
the above phase diagram.  However, the steady state found does depend
on the initial conditions used.  Specifically, considering small
periodic perturbations around the trivial state and time evolving
Eq.~(\ref{eq:5}) yields the AFM, trivial and FM states exactly as
discussed above. In contrast, if time evolved from a random initial
configuration, domains of FM/AFM can exist, separated by defect sites
(domain walls). The dynamics of such domain walls becomes frozen
within the mean-field numerics.  The absence of long-range order seen
in the full MPO numerics can be considered as the effect of a
superposition of many different configurations of domain walls.

\subsection{Correlations vs transverse field in the Ising limit}
\label{sec:open-syst-corr}

We now turn to the properties of quantum correlations at $\Delta=1$
(the Ising model).  For comparison, we summarize here the ground state
properties, as studied
in~\cite{Osterloh2002,Osborne2002,Vidal2003a}. In the Ising model
entanglement is short ranged: Only nearest and next-nearest
neighboring spins are entangled. The magnitude of the nearest
neighbor entanglement however shows critical scaling. At the critical
point $|g|=1$ Ref.~\cite{Osterloh2002} showed that $d \mathcal{C} / d
g$ (where $\mathcal{C}$ is concurrence, another measure of
entanglement) scaled as a power of the system size.  Consequently, the
peak value of $\mathcal{C}(g)$ actually occurs for $|g|>1$, rather
than at the critical point.  In the ground state, nearest neighbor
entanglement only vanished at $g \to 0, |g| \to
\infty$~\cite{Osterloh2002}.  Quantum discord for the same model was
studied in Ref.~\cite{Maziero2010}. Discord is not restricted to
nearest neighbors, and  is peaked near $|g|=1$.

\begin{figure}[htpb]
  \centering
  \includegraphics[width=3.2in]{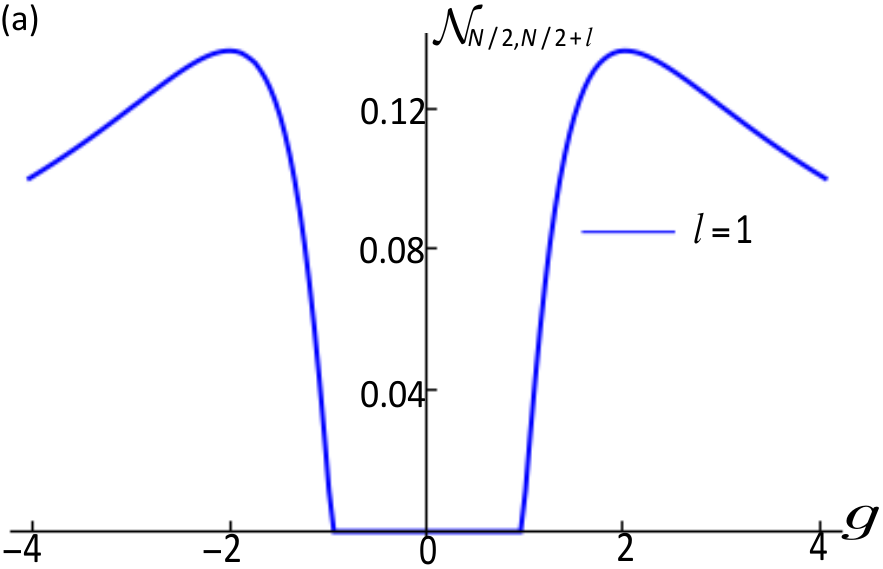}\\
    \includegraphics[width=3.2in]{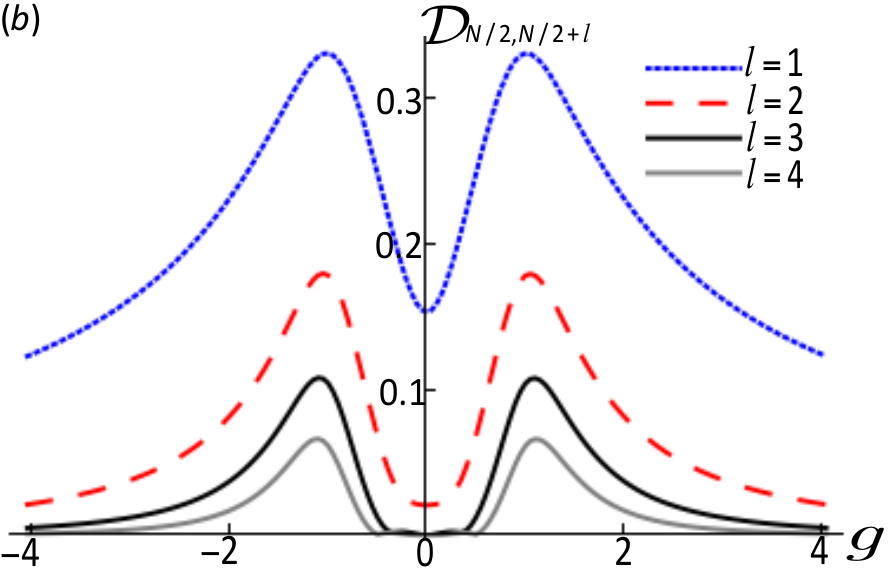}\\
     \includegraphics[width=3.2in]{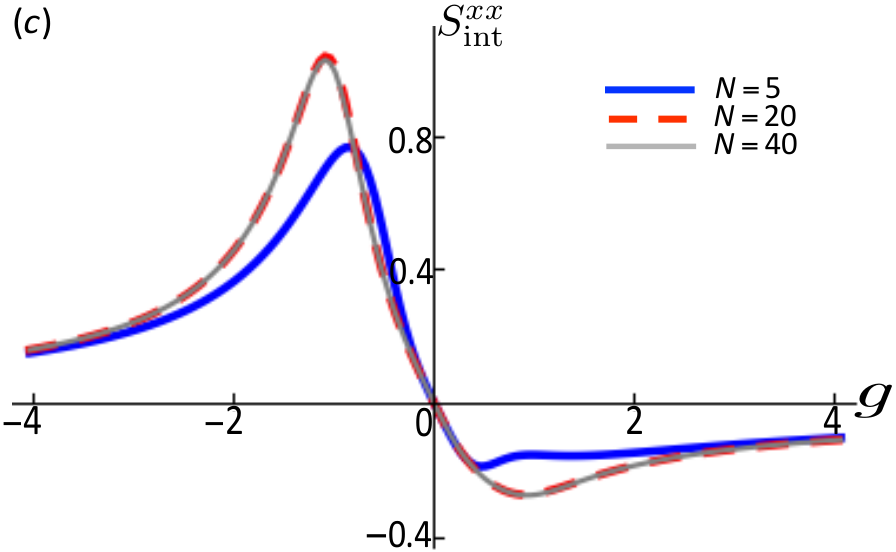}
     \caption{(Color online) Evolution of quantum correlations with
       transverse field $g$ in the Ising limit.  Panel (a) shows
       negativity $\mathcal N$, panel (b) geometric quantum discord
       $\mathcal D$.  In addition the integrated susceptibility
       $S^{xx}_{\text{int}} = \sum_j \langle \sigma_{i}^x \sigma_{j}^x
       \rangle$ is shown in (c); at the equilibrium critical point
       this would show a power law divergence with system size. Note that only one line is shown in panel (a) because
         entanglement vanishes beyond nearest neighbors at
         $\Delta=1$. Parameters as in Fig.~\ref{fig:isingorder}. }%
   \label{fig:ising}
\end{figure}

Figure~\ref{fig:ising} shows the evolution of quantum correlations
(negativity~\footnote{NB, the measure that we use, negativity, and
  that used in Refs.\cite{Osterloh2002,Osborne2002}, concurrence, are
  not identical.  We have checked that plotting concurrence instead of
  negativity does not affect any of the conclusions discussed here.}
and geometric quantum discord) with transverse field $g$ in the
non-equilibrium steady state at $\Delta=1, \kaptil=0.5$. In addition,
the integrated susceptibility $S^{xx}_{\text{int}} = \sum_j \langle
\sigma_{i}^x \sigma_{j}^x \rangle$ (static spin structure factor) is
shown.  This correlation function both serves as an example of a
correlation function that does not require specifically quantum
correlations, and also as a function which would diverge (as a power
law of system size) at the ground state critical point --- such a
divergence reflects the appearance of quasi-long range order in the
spin-spin correlator.  The asymmetry of this correlation function seen
in Fig.~\ref{fig:ising}(c) reflects the switch from ferromagnetic to
antiferromagnetic order.

Despite the switch between ferro- and antiferromagnetic order with
sign of $g$, which is absent in the ground state, several features of
the quantum correlations match closely the ground state behavior. Entanglement has a short range, existing now only between nearest
  neighbors as shown in Fig.~\ref{fig:ising}(a), while discord extends
  to greater separations Fig.~\ref{fig:ising}(b). Negativity also
peaks at a value $|g|>1$.  These features exist for both signs of $g$;
this is because the entanglement measures are not affected by the
sublattice sign-changes induced by the duality discussed above. As discussed in Ref.~\cite{Ma2011}, two-mode squeezing is a
  sufficient condition for pairwise entanglement.  We have confirmed
  that in the range of $g$ for which which bipartite entanglement
  vanishes, the two-mode spin squeezing parameter is identically zero.

In contrast to the ground state, there is however no critical
behavior: the entanglement is an analytic function of $g$ with no
singular behavior at $|g|=1$.  Similarly, the integrated
susceptibility does not diverge with increasing system size but
instead saturates. This reflects exponential spatial decay of
correlations, i.e. a finite correlation length, as anticipated due to
the dissipation.  The absence of critical behavior and the presence of
only short range correlations suggests the NESS of this 1D system does
not undergo any phase transition.  Such a result is to be expected,
since any finite temperature leads to short range order for a 1D
system with short ranged interactions. Although we consider
dissipation due to an empty (i.e. zero temperature) bath, we consider
a non-equilibrium situation. As has been discussed elsewhere, see
e.g.~\cite{Torre2011,Torre2012}, this leads to a non-zero low-energy
effective temperature.

Also in contrast to the ground state behavior, for small $|g|$
entanglement vanishes entirely.  The nature of this disappearance,
i.e.\ the sharp threshold seen in Fig.~\ref{fig:ising}(a) is a general
feature of entanglement in a dissipative system~\cite{Yu2009} ---
finite amounts of dissipation can make a state become separable.
Discord however remains non-zero between nearest neighbors at $g=0$.

\subsection{Correlations vs anisotropy (pump-strength) $\Delta$}
\label{sec:open-syst-corr-1}

In the ground state, the range of entanglement was found to grow as
one moves away from the Ising limit ($\Delta=1$), toward the isotropic
XY limit ($\Delta=0$)~\cite{Osborne2002}.  We therefore next explore
how pump strength $\Delta$ affects the scale and range of
correlations.  Since the anisotropy parameter $\Delta$ is also the
strength of pumping the isotropic limit corresponds to vanishing pump,
the consequence of this double r\^ole of $\Delta$.

\begin{figure}[htpb]
  \centering
  \includegraphics[width=3.2in]{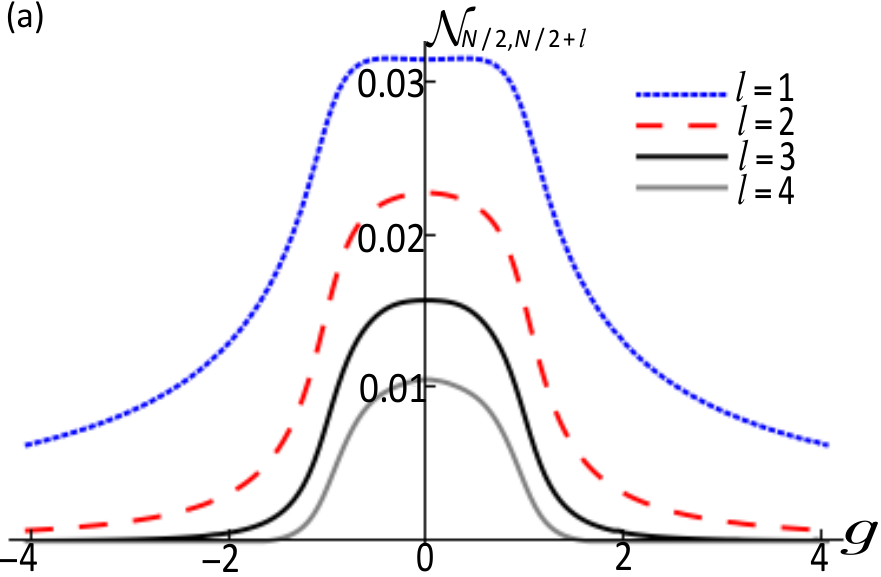}\\
    \includegraphics[width=3.2in]{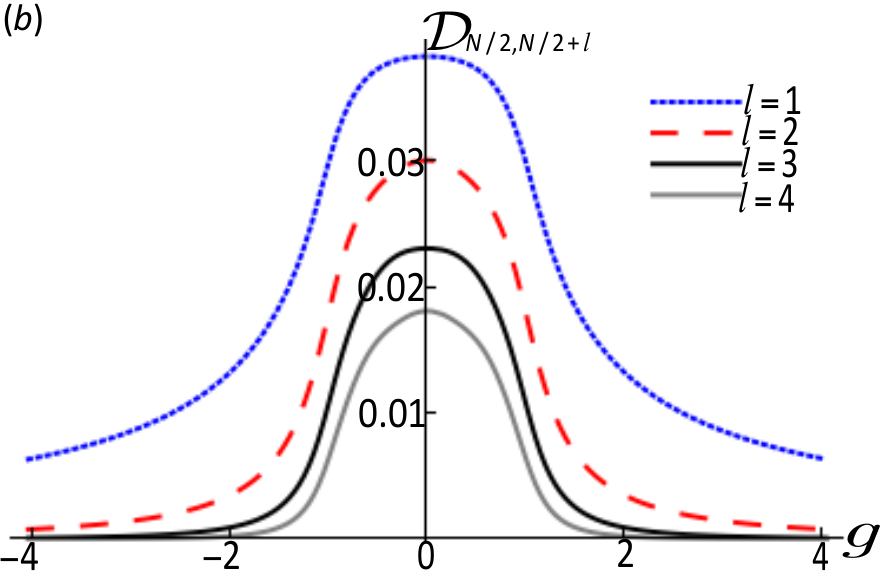}\\
     \includegraphics[width=3.2in]{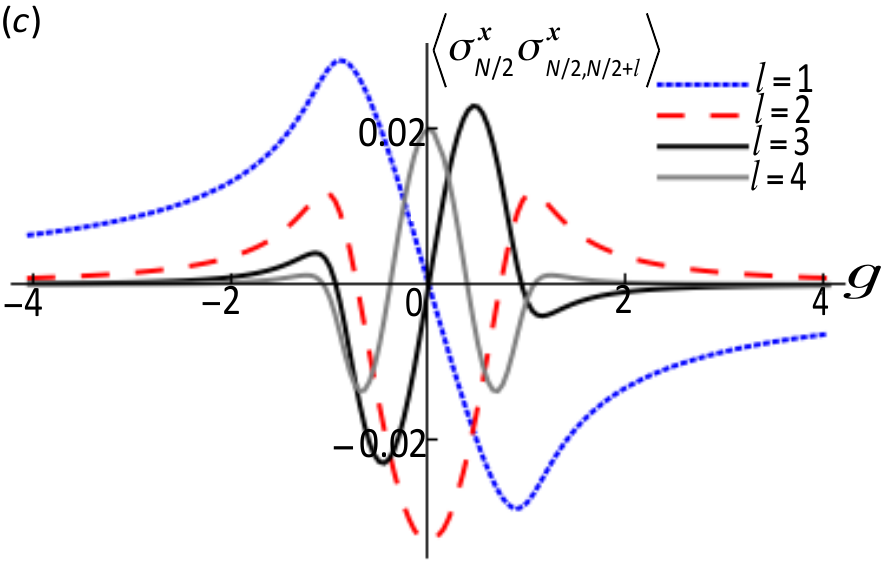}\\
     \includegraphics[width=3.2in]{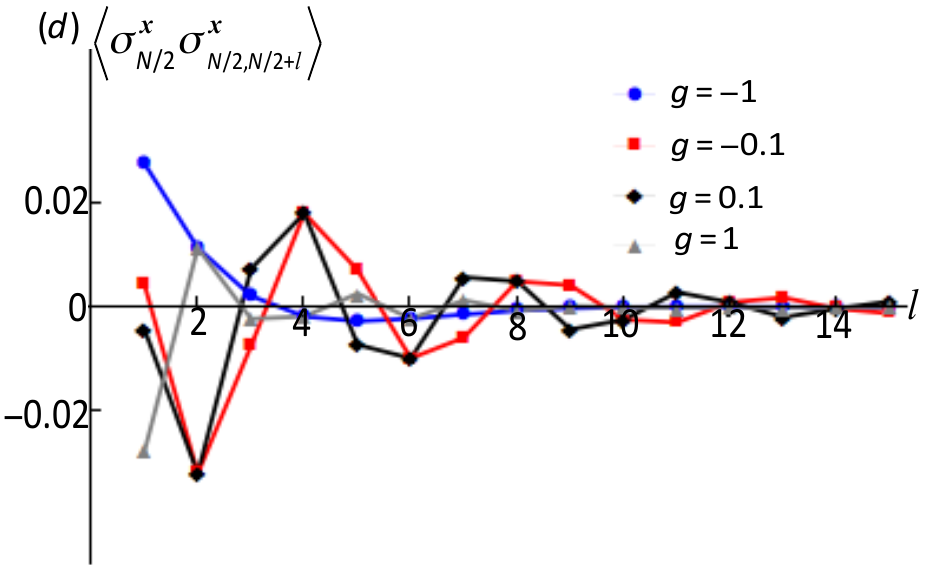}
     \caption{(Color online) Evolution of quantum correlations with
       transverse field $g$ near the isotropic limit, $\Delta=0.05$.
       Panels (a),(b) show negativity $\mathcal N$ and geometric
       quantum discord $\mathcal D$ as in Fig.~\ref{fig:ising}.  Panel
       (c) shows spin-spin correlation $\langle \sigma^x_{j}
       \sigma^x_{j+l} \rangle$ as in Fig.~\ref{fig:isingorder}. Panel
       (d) shows spatial dependence of correlations for the
       anisotropic X-Y model.  Parameters (in units of $J$):
       $\kaptil=0.5$ and MPO calculation performed for $N=40$ site
       chain, with $\chi_{\text{max}}=20$.  }%
     \label{fig:xy}
\end{figure}
 
We first consider how Fig.~\ref{fig:ising} is modified when
$\Delta<1$.  Figure~\ref{fig:xy} shows the behavior of entanglement,
discord and correlation functions for $\Delta=0.05$, close to the
isotropic limit.  As discussed above, the $\langle \sigma^x_{N/2}
\sigma^x_{N/2+l} \rangle$ still show the odd even symmetry, but the
vanishing of all correlations at $g=0$ no longer occurs --- the
precession axis now lies within the $xy$ plane, and so the $x$
component of spin need not decay to zero.  When $\Delta<1$, as in the
ground state, entanglement extends over a larger range, i.e.\ not only
between nearest neighbors.  In addition, the peak entanglement now
occurs near $g=0$, rather than at $|g|>1$, i.e.\ quantum correlations
attain their maximal value away from the equilibrium quantum critical
point~\cite{Latorre2004a}. In addition, the peak value of
  entanglement (and all correlations) is significantly smaller than
  that seen at $\Delta=1$. From Fig.~\ref{fig:xy}(c) it is clear that
  at large negative $g$ there is again short-range ferromagnetic
  order, and antiferromagnetic order at large positive $g$.  At smaller
  $g$, just as seen at $\Delta=1$, the short-range ordering is
  incommensurate [see Fig.~\ref{fig:xy}(d)].  However, the value of
  $|g|$ required to see FM/AFM order is larger for $\Delta=0.05$ than
  it was for $\Delta=1$, so that $g=\pm1$ now shows incommensurate
  order.The correlations do
  still respect the sublattice duality discussed earlier.  In contrast
  to the behavior at $\Delta=1$, the correlations always have a small
  magnitude [compare the scale of Fig.~\ref{fig:xy}(c,d) as compared
  to Fig.~\ref{fig:isingorder}].  This is consistent with the
  observation that for $\Delta=0.05, \tilde{\kappa}=0.5$, the
  mean-field theory would predict the trivial state independent of the
  value of $g$ (see Fig.~\ref{fig:meanfield}).

\begin{figure}[htpb]
  \centering
  \includegraphics[width=3.2in]{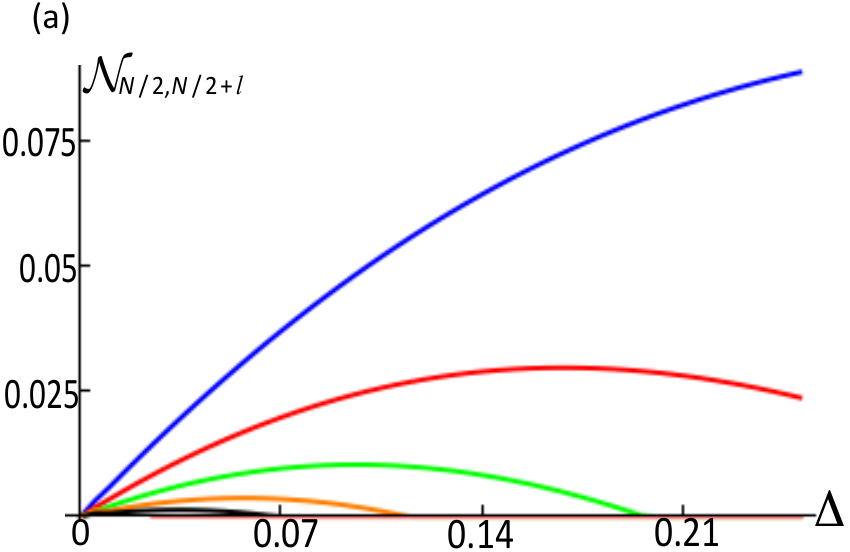}\\
    \includegraphics[width=3.2in]{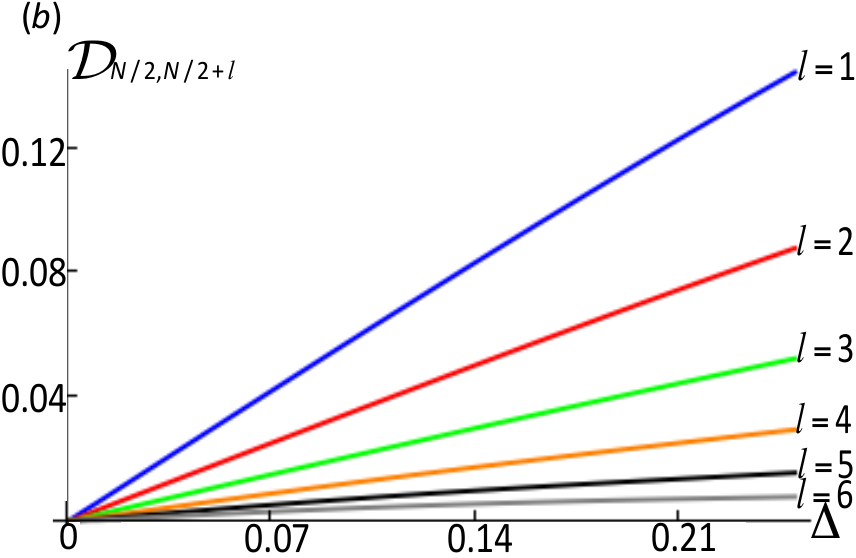}
    \caption{%
      (Color online) Evolution of quantum correlations with anisotropy
      $\Delta$. Panel (a) shows negativity $\mathcal N$ and (b)
      geometric quantum discord $\mathcal D$.  Parameters (in units of $J$): $g=-1,
      \kaptil=0.5$ and MPO calculation performed for $N=40$ site
      chain, with $\chi_{\text{max}}=20$.  }%
   \label{fig:vs-delta}
\end{figure}

While $\Delta=0.05$ leads to a longer range of entanglement, the
symmetry of the problem remains Ising-like for all $0<\Delta \le 1$.
In the ground state, the combination of this fact and universality
together imply that the range of entanglement must remain finite as
long as $\Delta$ is non-zero~\cite{Maziero2010}.  The same behavior
is indeed seen in the non-equilibrium steady state: For any non-zero
$\Delta$, entanglement only extends over a finite range, this range
grows as $\Delta$ shrinks and diverges at $\Delta \to 0$.  This can be
seen in Fig.~\ref{fig:vs-delta} which shows the evolution of
entanglement and discord as a function of $\Delta$ for various
different separations between sites.

As anticipated above, the limit $\Delta \to 0$ is special, since
$\Delta$ corresponds to pumping strength.  Specifically, as $\Delta
\to 0$, the range over which entanglement exists continues to grow,
but the magnitude of the entanglement for any pair of sites ultimately
vanishes.  Thus the limit $\Delta\to 0$ is singular, with diverging
range of correlations, but vanishing magnitude.  The vanishing of
negativity, and in fact of all correlations, at $\Delta=0$ can be
easily understood from the equation of motion: at $\Delta=0$, the
Hamiltonian conserves numbers of excited two-level systems, while the
dissipation reduces this number, so the steady state must be the
trivial empty state, which is a product state and thus uncorrelated.

The origin of growing range of negativity can be found by examining
the structure and scaling of the two-site density matrix.  We
first note that this density matrix has a simple structure:
\begin{equation}\label{eqnden}
\rho_{ij}=  \left(
    \begin{array}{cccc}
      p_{11}    & 0      & 0    & x_{4} \\
  0  & p_{10}   & x_{5}  & 0 \\
         0  & x_{5}^{\ast}    & p_{01}   & 0  \\
   x_{4}^{\ast}   & 0      & 0    & p_{00}.
  \end{array}\right).
\end{equation}
This structure is due to a symmetry of the equation of motion, under
the transformation $\rho \to \hat{R} \rho \hat{R}$ with $\hat{R} =
\prod_j \sigma^z_j$.  The consequences of such a symmetry for the
Hamiltonian were previously discussed~\cite{Osborne2002}; the decay
terms we consider also respect this symmetry.  Consequently the steady
state density matrix should satisfy $[R, \rho]=0$.  Tracing over all
but two sites, $[\sigma^z_{i}\sigma^z_{j},\rho_{ij}]=0$, which imposes
the structure discussed above.  A state of the form \eqref{eqnden} is
entangled iff either $p_{10}p_{01} <|x_4|^{2}$ or $p_{00} p_{11}
<|x_5|^{2}$.  In the limit of small $\Delta$ the excited state
populations $p_{11}, p_{01}, p_{10} \sim \Delta^2$ and so $p_{00} \sim
1$.  The off diagonal matrix elements scale as $|x_{4}| \sim \Delta,
|x_{5}| \sim \Delta^{2}$.  All of these expressions have prefactors
that depend on the separation between sites. However, regardless of
these prefactors, the scaling of $p_{01}, p_{10}, x_4$ with $\Delta$
implies that as $\Delta \to 0$, the first of the two criteria above
will always be satisfied,  i.e.\ for any pair of sites, there exists a
$\Delta_c$ such that for $0<\Delta<\Delta_c$ they will be entangled.
Furthermore, as discussed in section~\ref{sec:asymptotic-delta-to},
this behavior can be derived analytically within a spin-wave
approximation.

\subsection{Correlations vs decay rate}
\label{sec:corr-vs-decay}

\begin{figure}[htpb]
  \centering
  \includegraphics[width=3.2in]{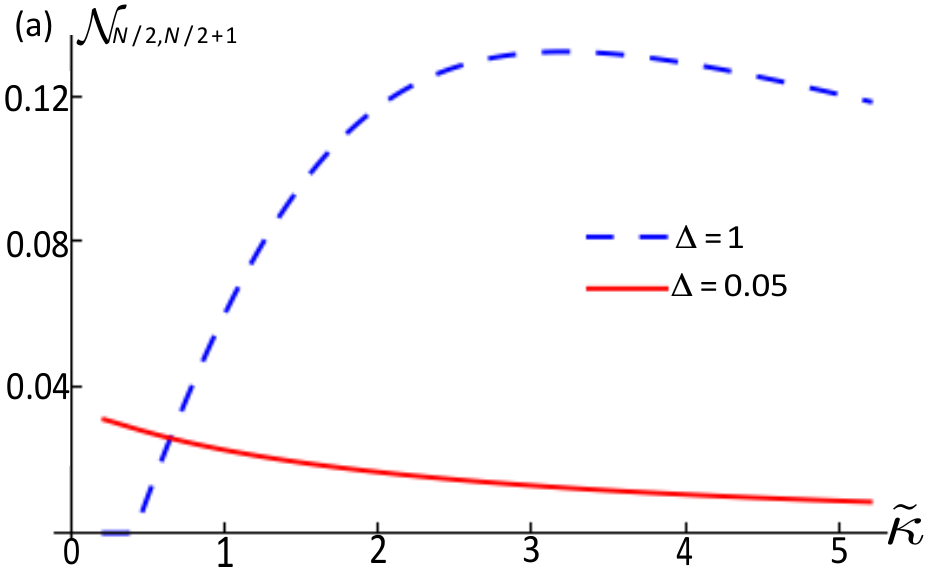}\\
    \includegraphics[width=3.2in]{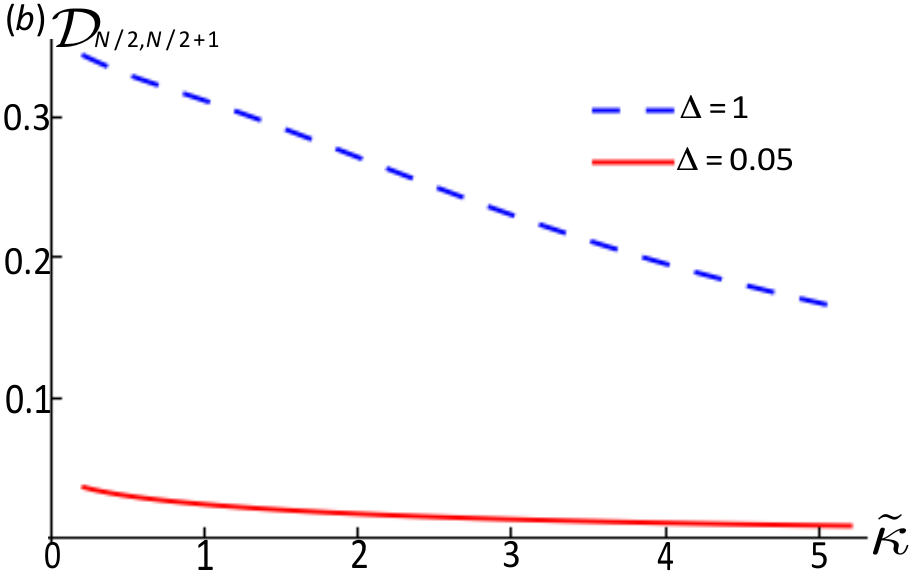}
    \caption{(Color online) Evolution of quantum correlations with
      $\kaptil$.  Panel (a) shows negativity $\mathcal N$, and panel
      (b) shows geometric quantum discord $\mathcal D$, for both Ising
      limit and small anisotropy limit. Parameters (in units of $J$): $g=-1$ and MPO
      calculation performed for $N=40$ site chain, with
      $\chi_{\text{max}}=20$.  }
       \label{fig:vs-decay}
\end{figure}

Having explored the dependence on the parameters $\Delta, g$, we
conclude our discussion of numerical results by presenting the
dependence of quantum correlations on the decay rate
$\kaptil=\kappa/J$.  Figure~\ref{fig:vs-decay} shows the evolution
with decay rate at $g=-1$, and the two values of $\Delta$ shown in
detail above.  Whereas the discord decreases monotonically with decay
rate, the behavior of the negativity depends on anisotropy.  In
particular, in the Ising limit, there is a non-monotonic dependence,
exhibiting a separable but non-classical state for sufficiently small
$\kaptil$.  The appearance of non-zero entanglement with increasing
$\kaptil$ corresponds to the condition $p_{01} p_{10}=|x_4|^2$: on
increasing $\kaptil$, the probabilities $p_{01}\equiv p_{10}$ decrease
while $|x_4|$ varies little at small $\kaptil$.  Non-monotonic
dependence of entanglement on decay rate has also been seen in other
contexts~\cite{Joshi2013}.  Note that the decay terms remain important
even at $\kaptil \to 0$. In this limit the steady state is only
attained at long times; the state which is finally attained is still
determined by the open system dynamics.

\section{Asymptotic $\Delta \to 0$ behavior and spin-wave approximation}
\label{sec:asymptotic-delta-to}
\subsection{Spin-wave calculation of negativity}
\label{sec:spin-wave-calc}

As noted above, for $\Delta=0$, the NESS of our model corresponds to
an empty state.  This suggests that for small $\Delta$ an
approximation based on a low density of excited two-level systems can
be used: a bosonic spin-wave approach~\cite{mattis2006theory}.  This
corresponds to reverting from two-level systems (hard core bosons) to
bosonic fields $\sigma_{j}^{-} \rightarrow
\hat{b}_{j}$. Equation~(\ref{eqnham2}) thus becomes:
\begin{multline}
  \label{eq:heff_boson}
  H_\text{eff} =- \sum_{j} \left[
    g(2\hat{b}_{j}^{\dagger}\hat{b}_{j}-1) 
    + \left(
      \hat{b}_{j}^{\dagger}\hat{b}_{j+1}+\hat{b}_{j+1}^{\dagger}\hat{b}_{j} 
    \right)
    \right.\\\left.
    {}+\Delta 
    (\hat{b}_{j}^{\dagger}\hat{b}_{j+1}^{\dagger}+\hat{b}_{j+1}\hat{b}_{j}) \right].
\end{multline}
This approximation is valid as long as double occupancy of a site can
be ignored.  Fourier transforming both this and the loss term, the
Master equation can be written as:
\begin{align}\label{harcorebosmas}
  \frac{d\rho}{dt}&=-i[ \sum_{k} h_k, \rho ]
  +
  \kaptil \!
  \sum_{k}
  [2 \hat{b}_{k}{\rho}\hat{b}_{k}^{\dagger}-\hat{b}_{k}^{\dagger}\hat{b}_{k}\rho-
  \rho\hat{b}_{k}^{\dagger}\hat{b}_{k}]
  \\
  h_{k}&=- 
  \left(\begin{array}{cc}
      \hat{b}^{\dagger}_{k}& \hat{b}_{-k}  
    \end{array}\right)
  \left(
    \begin{array}{cc}
      g + \cos(k) & \Delta \cos(k) \\
      \Delta \cos(k) & g + \cos(k)
    \end{array}
  \right)
  \left(
    \begin{array}{c}
      \hat{b}_k\\ \hat{b}^{\dagger}_{-k} 
    \end{array}
  \right)
  \nonumber,
\end{align}
so that each pair of modes $k, -k$ form a closed subsystem.

To find steady state correlations, we replace the density matrix
equation of motion, Eq.~(\ref{harcorebosmas}), by equivalent
Heisenberg-Langevin equations~\cite{scully97}.  The
Heisenberg-Langevin equations can be derived by writing the Heisenberg
equations for the system operators coupled to a Markovian bath.  After
eliminating the dynamics of the bath operators, one finds equations
for the system operators of the form:
\begin{equation}
\label{sec:spin-wave-calc-2}
\frac{d}{dt} \hat{b}_{k}=i[h_{k}+h_{-k},\hat{b}_{k}]-\kaptil \hat{b}_{k}+\sqrt{2 \kaptil} \hat{b}_{k}^{in}(t).
\end{equation}
The Markovian bath has two effects: it causes decay of the system
operator $\hat{b}_{k}$ at the rate $\kaptil$, and it introduces an
``input noise'' term $\hat{b}_{k}^{in}(t)$.  Since we consider decay
into an zero temperature (i.e. empty) bath, there is only vacuum
quantum noise: the only non-zero noise correlation function is
$\langle \hat{b}_{k}^{in}(t) \hat{b}_{k^{\prime}}^{\dagger
  in}(t^{\prime}) \rangle=\delta_{k,k^{\prime}}\delta(t-t^{\prime})$.
Because of the anomalous (pumping) terms in Eq.~(\ref{eq:heff_boson}),
the equation for $\hat{b}_k$ couples to that for $\hat{b}_{-k}^{\dagger}$ and
vice versa. The coupled equations for operators $\hat{b}_{k}$ and $\hat{b}_{-k}^{\dagger}$  can be written in a matrix form, 
\begin{align}\label{eqndif}
\dot{\hat{f}}(t)
  &=\mathcal M \hat{f}(t)+ \hat{m}(t), 
  \end{align}
    in which $\hat{f}(t)$ is the column vector comprising operators $\hat{b}_{k}(t)$ and $\hat{b}_{-k}^{\dagger}(t)$, $\hat{m}(t)$ is the column vector containing the noise operators:
    \begin{align}
\hat{f}(t)
  &=\! \left(\!
    \begin{array}{cc}
    \hat{b}_{k}(t),
      &
  \hat{b}_{-k}^{\dagger}(t)
    \end{array}\!\!\right)^T \nonumber \\
   \hat{m}(t)
  &=\! \left(\!
    \begin{array}{cc}
  \sqrt{2\kaptil}\hat{b}_{k}^{in}(t),
      &
\sqrt{2\kaptil}\hat{b}_{-k}^{\dagger in}(t)
    \end{array}\!\!\right)^T \nonumber;
  \end{align}
    and the  matrix $\mathcal M$ is given by 
     \begin{align}
 \mathcal{M}= \left(\!
    \begin{array}{cc}
     -\kaptil+ 2i(g+\cos(k))
      &
     2i\Delta\cos(k)
      \\
     -2i\Delta\cos(k)
      &
     -\kaptil-2i(g+\cos(k))
    \end{array}\!\!\right).\nonumber
  \end{align}
 The solution of Eq.~(\ref{eqndif}) is \begin{math} \hat{f}(t)=e^{\mathcal M t}\hat{f}(0)+\int_{0}^{t}e^{\mathcal M (t-t')}\hat{m}(t')dt' \end{math}. Since
the real parts of the eigenvalues of $\mathcal M$ are negative,
the first of these terms vanishes in the long time limit $t \to \infty$. 
In this limit one then finds:
\begin{align}
  \frac{\hat{b}_{k}(t)}{\sqrt{2 \kaptil}}
  &=\!
   \int_{0}^{t}
  \!\! dt^\prime \left[
    \mathcal G_{1} (t-t^\prime)\hat{b}_{k}^{in}(t^\prime)
    +\mathcal G_{2}(t-t^\prime) \hat{b}_{-k}^{\dagger in}(t^\prime) 
  \right]
  \nonumber\\
  \frac{\hat{b}_{-k}^{\dagger}(t)}{\sqrt{2 \kaptil}}
  &=\!
  \int_{0}^{t}
  \!\! dt^\prime \left[
    \mathcal G_{1}^{\ast} (t-t^\prime)\hat{b}_{-k}^{\dagger in}(t^\prime)
    +
    \mathcal G_{2}^{\ast}(t-t^\prime) \hat{b}_{k}^{n}(t^\prime))
  \right]
  \label{eq:3}
\end{align}
where the propagators $\mathcal G_{1,2}(\tau)$ are matrix
elements of $\exp(M \tau)$. By introducing the
dispersions
\begin{math}
 \epsilon_{k}=2(g+\cos(k)), \eta_{k}=2\Delta\cos(k),
  \xi_k =
  \sqrt{\epsilon_k^2 - \eta_k^2},
\end{math}
the propagators can be written as:
\begin{align*}
\mathcal G_{1}(\tau)&=e^{-\kaptil \tau} 
\left[
  \cos(\tau \xi_k)+i \epsilon_{k} \frac{\sin \left(\tau \xi_k\right)}{\xi_k}
\right]\\
\mathcal G_{2}(\tau)&=
i \eta_{k} e^{-\kaptil \tau} \sin(\tau \xi_k)/\xi_k.
\end{align*}

To find the quantum correlations of the state, we  first note that
since the problem involves non-interacting bosons, the steady state is
Gaussian, i.e.\ it can be fully characterized by the covariance matrix
$V_{j,k}$ as given below.  Introducing $\hat{x}_j=\hat{b}^{}_j +
\hat{b}^\dagger_j, \hat{p}_j =(\hat{b}^{}_j - \hat{b}^\dagger_j)/i$ we have
\begin{equation}\label{eqndcov}
  V_{j,k}=  \!\left(\!
    \begin{array}{cc}
      {\bf A}_{j}    & {\bf C}_{jk}\\
      {\bf C}_{jk}^{T} & {\bf A}_{k} 
    \end{array}\!\!\right)\!,
  \quad
  {\bf C}_{jk}=\!  \left(\!
    \begin{array}{cc}
      \langle x_j x_k \rangle_s
      &
      \langle x_j p_k \rangle_s
      \\
      \langle x_k p_j \rangle_s
      &
      \langle p_j p_k \rangle_s
    \end{array}\!\!\right),
\end{equation}
and ${\bf A}_j = {\bf C}_{jj}$ where $\langle xp\rangle_s = \langle xp
+px\rangle/2$. To find these correlators, it is sufficient to
find $\langle \hat{b}_{j}^{\dagger}\hat{b}_{j+l} \rangle$, and $\langle
\hat{b}_{j}\hat{b}_{j+l} \rangle$. In the real space the correlator $\langle \hat{b}_{j}^{\dagger}\hat{b}_{j+l} \rangle$ can be expressed as
\begin{equation}\label{realspce}
 \langle \hat{b}_{j}^{\dagger} \hat{b}^{}_{j+l} \rangle
 =\frac{1}{N} \sum_{k,k'} \langle \hat{b}_{k}^{\dagger} \hat{b}^{}_{k'}
 \rangle e^{i(k'-k)j}e^{ik' l} .
\end{equation}
Using Eq.~(\ref{eq:3}) one finds that for $N \to \infty$:
\begin{equation}\label{ctrint}
 \langle \hat{b}_{j}^{\dagger} \hat{b}_{j+l} \rangle
 =
 \frac{1}{4\pi}\int_{-\pi}^{\pi} \frac{\eta_k^2}{\xi_k^2+\kaptil^{2}}
 dk e^{i k l}
\end{equation}
and a similar expression for $ \langle \hat{b}_{j}^{} \hat{b}_{j+l}
\rangle$. By substituting $e^{ik} \to z$, the integral becomes a
contour integral around the unit circle $|z|=1$, so its value depends
on the residue of those poles $z=Z$ with $ |Z|<1$.  The four poles
come in complex conjugate pairs and can be found in closed form
$Z=\zeta\pm \sqrt{\zeta^2-1}$ where $\zeta=\left[ g \pm \sqrt{g^2
    \Delta^2 - \kaptil^2(1-\Delta^2)/4} \right]/(1-\Delta^2)$.  Two of
these poles which we denote as $Z_0, Z_0^\ast$ lie within the unit
circle, and in terms of these one finds:
\begin{align}
  \langle \hat{b}_{j}^{\dagger} \hat{b}_{j+l} \rangle &=
  \Delta^{2}
  \left[\alpha (Z_0)^{l-1}+\alpha^{*} (Z_0^{*})^{l-1} \right] \\
  \langle \hat{b}_{j} \hat{b}_{j+l} \rangle &=\Delta 
  \beta
  (Z_0^{*})^{l-1}
\end{align}
where $\alpha,\beta$ are complex functions of $\kaptil,\Delta,g$.  We
have factored out the asymptotic scaling with $\Delta$ at $\Delta \to
0$. Since $|Z_0|<1$, all correlations decay exponentially with
separation $n$.

The definition of negativity given earlier, Eq.~(\ref{eq:4}), is
specific to qubits, i.e.\ two-level systems.  For a Gaussian state an
alternate definition of negativity can be found in terms of the
symplectic eigenvalue $\tilde{\nu}_{-}^{2}=(\tau-\sqrt{\tau^{2}-4
  {\rm Det}[V_{j,j+l}]})/2$, where $\tau={\rm Det}[A_{j}]+{\rm
  Det}[A_{j+l}]-2{\rm Det}[C_{j,j+l}]$.  The state is separable if
$\tilde{\nu}_{-} > 1$, and so negativity for such states may be
defined as $\mathcal N=\text{max}( 0, 1-\tilde{\nu}_{-})$.  Using the
asymptotic scaling of the elements of the covariance matrix with $\Delta$, we
find that in the $\Delta \to 0$ limit
\begin{equation}\label{eqn:nega}
  \tilde{\nu}_{-} \simeq\sqrt{1-4|\langle \hat{b}_{j}  \hat{b}_{j+l} \rangle|^2},
  \quad
  \mathcal N \simeq 2|\langle \hat{b}_{j}  \hat{b}_{j+l} \rangle|
\end{equation}
Within this limit, it is thus clear that $\mathcal N > 0$ for all
pairs of sites, but $\mathcal N \propto \Delta$ and so $\mathcal N$
vanishes at small $\Delta$, reproducing the singular behavior found
numerically in the previous section.

\subsection{Comparing spin-wave approximation to numerics}
\label{sec:validity-spin-wave}

The spin wave theory relies on neglecting effects of possible double
occupation of a given site.  While the probability of such an event is
small for $\Delta \to 0$, it is not a-priori clear whether its effects
are negligible, since the pair creation term creates excitations on
adjacent sites, hence hopping can easily create a doubly occupied site
within the bosonic approximation.  For this reason, we a compare the
results of the MPO numerics and the spin-wave theory in the limit
$\Delta \to 0$.

\begin{figure}[htpb]
     \begin{center}
       \includegraphics[width=3.2in]{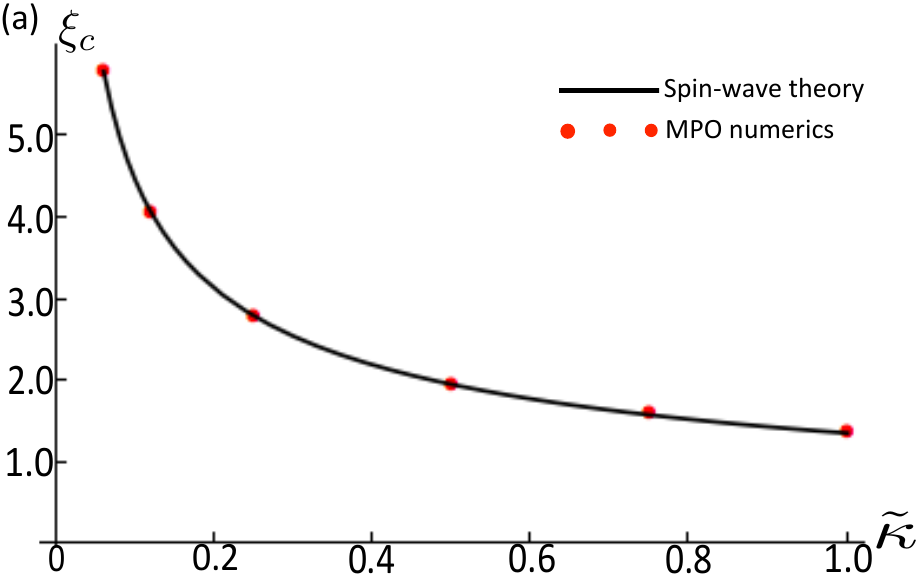}\\
       \includegraphics[width=3.2in]{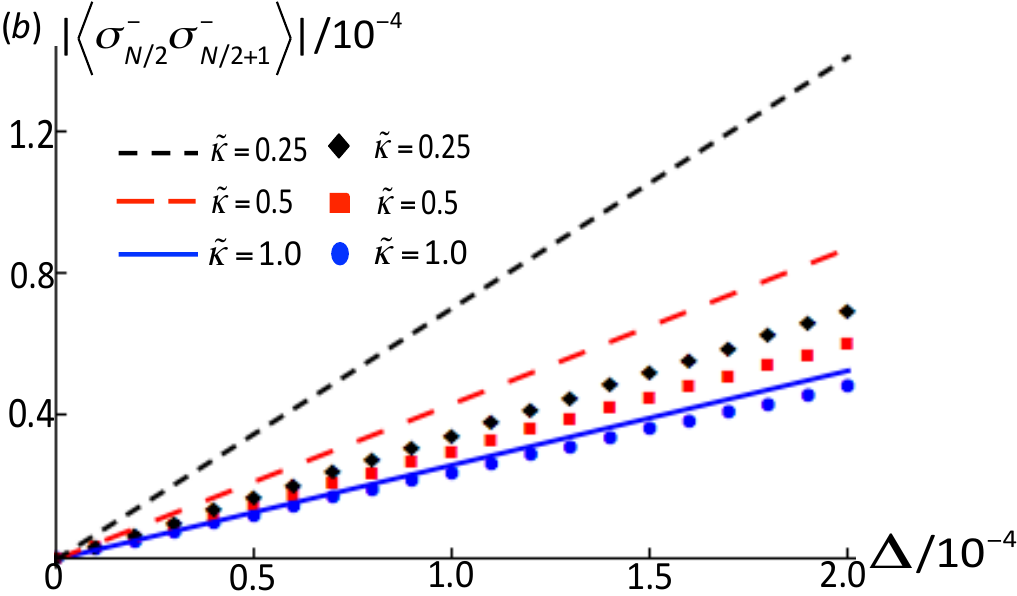}\\
    \end{center}
    \caption{%
      (Color online) Panel (a) correlation length $\xi_{c}$ at $\Delta
      = 0.005$ vs decay rate $\kaptil$.  Panel (b) Nearest neighbor
      correlations $|\langle\sigma_{j}^{-}\sigma_{j+1}^{-} \rangle|$
      vs anisotropy $\Delta$ for several values of $\kaptil$. In both
      panels MPO numerics (points) are compared to spin-wave theory
      (lines).  Parameters (in units of $J$): $g=-1$, MPO calculation is performed for
      $N = 40$ site chain with $\chi_{\text{max}}=20$.  }
   \label{fig:compare-mps-sw}
\end{figure}

We focus on the correlation function
$\langle\sigma_{j}^{-}\sigma_{j+l}^{-} \rangle$, or its equivalent
bosonic form, which according to Eq.~(\ref{eqn:nega}) determines the
asymptotic negativity as $\Delta \to 0$.  Both MPO and spin-wave
results show this correlation function decays exponentially with
separation $l$ (neglecting edge effects).  Consequently this
correlation function can be characterized by its value for nearest
neighbors $l=1$ (NB the $l=0$ case vanishes by definition), and by its
correlation length $\xi_c$, defined as
$|\langle\sigma_{j}^{-}\sigma_{j+l}^{-} \rangle| \propto e^{-l/\xi_c}$.
In the spin-wave theory $\xi_c=-1/\ln[Z_0]$.  These two characteristic
quantities are shown in Fig.~\ref{fig:compare-mps-sw}, focussing on
the limiting behavior at $\Delta \to 0$.

The correlation length shown in Fig.~\ref{fig:compare-mps-sw} shows
that the spin-wave theory accurately reproduces the results of the
numerics, and both show a diverging correlation length ($|Z_0| \to 1$)
in the limit $\kaptil\to 0$.  In contrast, the magnitude of
correlations (i.e.\ prefactors of the exponential decay) do not match
well except at $\kaptil \gg 1$.  This can be explained as follows: at
small $\kaptil$, excitations created on adjacent sites can easily hop
to create doubly occupied sites and thus rendering the bosonic
approximation inaccurate. For $\kaptil \gg 1$, excitations on adjacent
sites are lost before hopping can create doubly excited sites.

\section{Conclusions}
\label{sec:conclusions}
In the present work we have studied the non-equilibrium steady state
of parametrically driven 1-D coupled cavity array.  Making use of an
MPO representation to determine the open system evolution, we obtain
the non-equilibrium steady state of a dissipative transverse field
Ising model.  The steady state can be related to the ground state
configuration for transverse field $g<0$, and to the maximum energy
configuration for $g>0$.  Consequently, for either sign of $g$, many
features of the quantum correlations behave similarly to those in the
ground state Ising model.  The most significant difference is that
dissipation destroys the phase transition, and so no critical behavior
occurs at $|g|=1$ with correlation lengths remaining finite. We
  have also compared the results of the MPO numerics with the
  predictions of the mean-field theory. Mean-field theory erroneously
  predicts long-range ordered phases, but the nature of the ordering
  predicted is reflected by the MPO numerics. We have identified a
singular limit, of weak driving, where the range of quantum
correlations diverges, but the magnitude of the correlations
vanishes. This limiting behavior can be recovered analytically from a
spin wave theory, which accurately recovers the correlation length in
this limit.

\begin{acknowledgments}
  CJ and JK acknowledge support from EPSRC programme ``TOPNES''
  (EP/I031014/1), and EPSRC (EP/G004714/2). FN acknowledges support from the EPSRC  grant ``A Pragmatic Approach to Adiabatic Quantum Computation'' (EP/K02163X/1). JK acknowledges helpful
  suggestsions from R. Fazio and hospitality from MPI-PKS, Dresden.
  We acknowledge discussions with A.~G.~Green, C.~A.~Hooley and
  S.~H.~Simon.
\end{acknowledgments}

%\bibliography{IsingMPS}

\begin{thebibliography}{74}%
\makeatletter
\providecommand \@ifxundefined [1]{%
 \@ifx{#1\undefined}
}%
\providecommand \@ifnum [1]{%
 \ifnum #1\expandafter \@firstoftwo
 \else \expandafter \@secondoftwo
 \fi
}%
\providecommand \@ifx [1]{%
 \ifx #1\expandafter \@firstoftwo
 \else \expandafter \@secondoftwo
 \fi
}%
\providecommand \natexlab [1]{#1}%
\providecommand \enquote  [1]{``#1''}%
\providecommand \bibnamefont  [1]{#1}%
\providecommand \bibfnamefont [1]{#1}%
\providecommand \citenamefont [1]{#1}%
\providecommand \href@noop [0]{\@secondoftwo}%
\providecommand \href [0]{\begingroup \@sanitize@url \@href}%
\providecommand \@href[1]{\@@startlink{#1}\@@href}%
\providecommand \@@href[1]{\endgroup#1\@@endlink}%
\providecommand \@sanitize@url [0]{\catcode `\\12\catcode `\$12\catcode
  `\&12\catcode `\#12\catcode `\^12\catcode `\_12\catcode `\%12\relax}%
\providecommand \@@startlink[1]{}%
\providecommand \@@endlink[0]{}%
\providecommand \url  [0]{\begingroup\@sanitize@url \@url }%
\providecommand \@url [1]{\endgroup\@href {#1}{\urlprefix }}%
\providecommand \urlprefix  [0]{URL }%
\providecommand \Eprint [0]{\href }%
\providecommand \doibase [0]{http://dx.doi.org/}%
\providecommand \selectlanguage [0]{\@gobble}%
\providecommand \bibinfo  [0]{\@secondoftwo}%
\providecommand \bibfield  [0]{\@secondoftwo}%
\providecommand \translation [1]{[#1]}%
\providecommand \BibitemOpen [0]{}%
\providecommand \bibitemStop [0]{}%
\providecommand \bibitemNoStop [0]{.\EOS\space}%
\providecommand \EOS [0]{\spacefactor3000\relax}%
\providecommand \BibitemShut  [1]{\csname bibitem#1\endcsname}%
\let\auto@bib@innerbib\@empty
%</preamble>
\bibitem [{\citenamefont {Sachdev}(2011)}]{Sachdev2011}%
  \BibitemOpen
  \bibfield  {author} {\bibinfo {author} {\bibfnamefont {S.}~\bibnamefont
  {Sachdev}},\ }\href@noop {} {\emph {\bibinfo {title} {{Quantum Phase
  Transitions}}}},\ \bibinfo {edition} {2nd}\ ed.\ (\bibinfo  {publisher}
  {Cambridge University Press},\ \bibinfo {address} {Cambridge},\ \bibinfo
  {year} {2011})\BibitemShut {NoStop}%
\bibitem [{\citenamefont {Kadanoff}(2000)}]{kadanoff2000statistical}%
  \BibitemOpen
  \bibfield  {author} {\bibinfo {author} {\bibfnamefont {L.~P.}\ \bibnamefont
  {Kadanoff}},\ }\href {http://books.google.co.uk/books?id=6bKmQgAACAAJ} {\emph
  {\bibinfo {title} {{Statistical physics: statistics, dynamics and
  renormalization}}}},\ Statistical Physics: Statics, Dynamics and
  Renormalization\ (\bibinfo  {publisher} {World Scientific},\ \bibinfo
  {address} {Singapore},\ \bibinfo {year} {2000})\BibitemShut {NoStop}%
\bibitem [{\citenamefont {Nielsen}\ and\ \citenamefont
  {Chuang}(2010)}]{nielsen2010quantum}%
  \BibitemOpen
  \bibfield  {author} {\bibinfo {author} {\bibfnamefont {M.~A.}\ \bibnamefont
  {Nielsen}}\ and\ \bibinfo {author} {\bibfnamefont {I.~L.}\ \bibnamefont
  {Chuang}},\ }\href {http://books.google.co.uk/books?id=-s4DEy7o-a0C} {\emph
  {\bibinfo {title} {{Quantum Computation and Quantum Information}}}}\
  (\bibinfo  {publisher} {Cambridge University Press},\ \bibinfo {address}
  {Cambridge},\ \bibinfo {year} {2010})\BibitemShut {NoStop}%
\bibitem [{\citenamefont {Osterloh}\ \emph {et~al.}(2002)\citenamefont
  {Osterloh}, \citenamefont {Amico}, \citenamefont {Falci},\ and\ \citenamefont
  {Fazio}}]{Osterloh2002}%
  \BibitemOpen
  \bibfield  {author} {\bibinfo {author} {\bibfnamefont {A.}~\bibnamefont
  {Osterloh}}, \bibinfo {author} {\bibfnamefont {L.}~\bibnamefont {Amico}},
  \bibinfo {author} {\bibfnamefont {G.}~\bibnamefont {Falci}}, \ and\ \bibinfo
  {author} {\bibfnamefont {R.}~\bibnamefont {Fazio}},\ }\href {\doibase
  10.1038/416608a} {\bibfield  {journal} {\bibinfo  {journal} {Nature}\
  }\textbf {\bibinfo {volume} {416}},\ \bibinfo {pages} {608} (\bibinfo {year}
  {2002})}\BibitemShut {NoStop}%
\bibitem [{\citenamefont {Osborne}\ and\ \citenamefont
  {Nielsen}(2002)}]{Osborne2002}%
  \BibitemOpen
  \bibfield  {author} {\bibinfo {author} {\bibfnamefont {T.}~\bibnamefont
  {Osborne}}\ and\ \bibinfo {author} {\bibfnamefont {M.}~\bibnamefont
  {Nielsen}},\ }\href {\doibase 10.1103/PhysRevA.66.032110} {\bibfield
  {journal} {\bibinfo  {journal} {Phys. Rev. A.}\ }\textbf {\bibinfo {volume}
  {66}},\ \bibinfo {pages} {032110} (\bibinfo {year} {2002})}\BibitemShut
  {NoStop}%
\bibitem [{\citenamefont {Vidal}\ \emph {et~al.}(2003)\citenamefont {Vidal},
  \citenamefont {Latorre}, \citenamefont {Rico},\ and\ \citenamefont
  {Kitaev}}]{Vidal2003a}%
  \BibitemOpen
  \bibfield  {author} {\bibinfo {author} {\bibfnamefont {G.}~\bibnamefont
  {Vidal}}, \bibinfo {author} {\bibfnamefont {J.~I.}\ \bibnamefont {Latorre}},
  \bibinfo {author} {\bibfnamefont {E.}~\bibnamefont {Rico}}, \ and\ \bibinfo
  {author} {\bibfnamefont {A.}~\bibnamefont {Kitaev}},\ }\href {\doibase
  10.1103/PhysRevLett.90.227902} {\bibfield  {journal} {\bibinfo  {journal}
  {Phys. Rev. Lett.}\ }\textbf {\bibinfo {volume} {90}},\ \bibinfo {pages}
  {227902} (\bibinfo {year} {2003})}\BibitemShut {NoStop}%
\bibitem [{\citenamefont {Amico}\ \emph {et~al.}(2008)\citenamefont {Amico},
  \citenamefont {Osterloh},\ and\ \citenamefont {Vedral}}]{Amico2008}%
  \BibitemOpen
  \bibfield  {author} {\bibinfo {author} {\bibfnamefont {L.}~\bibnamefont
  {Amico}}, \bibinfo {author} {\bibfnamefont {A.}~\bibnamefont {Osterloh}}, \
  and\ \bibinfo {author} {\bibfnamefont {V.}~\bibnamefont {Vedral}},\ }\href
  {\doibase 10.1103/RevModPhys.80.517} {\bibfield  {journal} {\bibinfo
  {journal} {Rev. Mod. Phys.}\ }\textbf {\bibinfo {volume} {80}},\ \bibinfo
  {pages} {517} (\bibinfo {year} {2008})}\BibitemShut {NoStop}%
\bibitem [{\citenamefont {Horodecki}\ \emph {et~al.}(2009)\citenamefont
  {Horodecki}, \citenamefont {Horodecki}, \citenamefont {Horodecki},\ and\
  \citenamefont {Horodecki}}]{Horodecki2009a}%
  \BibitemOpen
  \bibfield  {author} {\bibinfo {author} {\bibfnamefont {R.}~\bibnamefont
  {Horodecki}}, \bibinfo {author} {\bibfnamefont {P.}~\bibnamefont
  {Horodecki}}, \bibinfo {author} {\bibfnamefont {M.}~\bibnamefont
  {Horodecki}}, \ and\ \bibinfo {author} {\bibfnamefont {K.}~\bibnamefont
  {Horodecki}},\ }\href {http://link.aps.org/doi/10.1103/RevModPhys.81.865}
  {\bibfield  {journal} {\bibinfo  {journal} {Rev. Mod. Phys.}\ }\textbf
  {\bibinfo {volume} {81}},\ \bibinfo {pages} {865} (\bibinfo {year}
  {2009})}\BibitemShut {NoStop}%
\bibitem [{\citenamefont {Breuer}\ and\ \citenamefont
  {Petruccione}(2002)}]{Breuer2002}%
  \BibitemOpen
  \bibfield  {author} {\bibinfo {author} {\bibfnamefont {H.-P.}\ \bibnamefont
  {Breuer}}\ and\ \bibinfo {author} {\bibfnamefont {F.}~\bibnamefont
  {Petruccione}},\ }\href@noop {} {\emph {\bibinfo {title} {{The Theory of Open
  Quantum Systems}}}}\ (\bibinfo  {publisher} {Oxford University Press},\
  \bibinfo {address} {Oxford},\ \bibinfo {year} {2002})\BibitemShut {NoStop}%
\bibitem [{\citenamefont {Zurek}(2003)}]{Zurek2003}%
  \BibitemOpen
  \bibfield  {author} {\bibinfo {author} {\bibfnamefont {W.~H.}\ \bibnamefont
  {Zurek}},\ }\href {\doibase 10.1103/RevModPhys.75.715} {\bibfield  {journal}
  {\bibinfo  {journal} {Rev. Mod. Phys.}\ }\textbf {\bibinfo {volume} {75}},\
  \bibinfo {pages} {715} (\bibinfo {year} {2003})}\BibitemShut {NoStop}%
\bibitem [{\citenamefont {{Lo Franco}}\ \emph {et~al.}(2013)\citenamefont {{Lo
  Franco}}, \citenamefont {Bellomo}, \citenamefont {Maniscalco},\ and\
  \citenamefont {Compagno}}]{LoFranco2013}%
  \BibitemOpen
  \bibfield  {author} {\bibinfo {author} {\bibfnamefont {R.}~\bibnamefont {{Lo
  Franco}}}, \bibinfo {author} {\bibfnamefont {B.}~\bibnamefont {Bellomo}},
  \bibinfo {author} {\bibfnamefont {S.}~\bibnamefont {Maniscalco}}, \ and\
  \bibinfo {author} {\bibfnamefont {G.}~\bibnamefont {Compagno}},\ }\href
  {http://www.worldscientific.com/doi/abs/10.1142/S0217979213450537} {\bibfield
   {journal} {\bibinfo  {journal} {Int. J. Mod. Phys. B}\ }\textbf {\bibinfo
  {volume} {27}},\ \bibinfo {pages} {1345053} (\bibinfo {year}
  {2013})}\BibitemShut {NoStop}%
\bibitem [{\citenamefont {Dimer}\ \emph {et~al.}(2007)\citenamefont {Dimer},
  \citenamefont {Estienne}, \citenamefont {Parkins},\ and\ \citenamefont
  {Carmichael}}]{Dimer2007}%
  \BibitemOpen
  \bibfield  {author} {\bibinfo {author} {\bibfnamefont {F.}~\bibnamefont
  {Dimer}}, \bibinfo {author} {\bibfnamefont {B.}~\bibnamefont {Estienne}},
  \bibinfo {author} {\bibfnamefont {A.~S.}\ \bibnamefont {Parkins}}, \ and\
  \bibinfo {author} {\bibfnamefont {H.~J.}\ \bibnamefont {Carmichael}},\ }\href
  {\doibase 10.1103/PhysRevA.75.013804} {\bibfield  {journal} {\bibinfo
  {journal} {Phys. Rev. A}\ }\textbf {\bibinfo {volume} {75}},\ \bibinfo
  {pages} {013804} (\bibinfo {year} {2007})}\BibitemShut {NoStop}%
\bibitem [{\citenamefont {Hartmann}(2010)}]{Hartmann2010a}%
  \BibitemOpen
  \bibfield  {author} {\bibinfo {author} {\bibfnamefont {M.~J.}\ \bibnamefont
  {Hartmann}},\ }\href {\doibase 10.1103/PhysRevLett.104.113601} {\bibfield
  {journal} {\bibinfo  {journal} {Phys. Rev. Lett.}\ }\textbf {\bibinfo
  {volume} {104}},\ \bibinfo {pages} {113601} (\bibinfo {year}
  {2010})}\BibitemShut {NoStop}%
\bibitem [{\citenamefont {Baumann}\ \emph {et~al.}(2010)\citenamefont
  {Baumann}, \citenamefont {Guerlin}, \citenamefont {Brennecke},\ and\
  \citenamefont {Esslinger}}]{Baumann2010}%
  \BibitemOpen
  \bibfield  {author} {\bibinfo {author} {\bibfnamefont {K.}~\bibnamefont
  {Baumann}}, \bibinfo {author} {\bibfnamefont {C.}~\bibnamefont {Guerlin}},
  \bibinfo {author} {\bibfnamefont {F.}~\bibnamefont {Brennecke}}, \ and\
  \bibinfo {author} {\bibfnamefont {T.}~\bibnamefont {Esslinger}},\ }\href
  {\doibase 10.1038/nature09009} {\bibfield  {journal} {\bibinfo  {journal}
  {Nature}\ }\textbf {\bibinfo {volume} {464}},\ \bibinfo {pages} {1301}
  (\bibinfo {year} {2010})}\BibitemShut {NoStop}%
\bibitem [{\citenamefont {Nagy}\ \emph {et~al.}(2010)\citenamefont {Nagy},
  \citenamefont {K\'{o}nya}, \citenamefont {Szirmai},\ and\ \citenamefont
  {Domokos}}]{Nagy2010}%
  \BibitemOpen
  \bibfield  {author} {\bibinfo {author} {\bibfnamefont {D.}~\bibnamefont
  {Nagy}}, \bibinfo {author} {\bibfnamefont {G.}~\bibnamefont {K\'{o}nya}},
  \bibinfo {author} {\bibfnamefont {G.}~\bibnamefont {Szirmai}}, \ and\
  \bibinfo {author} {\bibfnamefont {P.}~\bibnamefont {Domokos}},\ }\href
  {\doibase 10.1103/PhysRevLett.104.130401} {\bibfield  {journal} {\bibinfo
  {journal} {Phys. Rev. Lett.}\ }\textbf {\bibinfo {volume} {104}},\ \bibinfo
  {pages} {1} (\bibinfo {year} {2010})}\BibitemShut {NoStop}%
\bibitem [{\citenamefont {Diehl}\ \emph {et~al.}(2010)\citenamefont {Diehl},
  \citenamefont {Tomadin}, \citenamefont {Micheli}, \citenamefont {Fazio},\
  and\ \citenamefont {Zoller}}]{Diehl2010}%
  \BibitemOpen
  \bibfield  {author} {\bibinfo {author} {\bibfnamefont {S.}~\bibnamefont
  {Diehl}}, \bibinfo {author} {\bibfnamefont {A.}~\bibnamefont {Tomadin}},
  \bibinfo {author} {\bibfnamefont {A.}~\bibnamefont {Micheli}}, \bibinfo
  {author} {\bibfnamefont {R.}~\bibnamefont {Fazio}}, \ and\ \bibinfo {author}
  {\bibfnamefont {P.}~\bibnamefont {Zoller}},\ }\href {\doibase
  10.1103/PhysRevLett.105.015702} {\bibfield  {journal} {\bibinfo  {journal}
  {Phys. Rev. Lett.}\ }\textbf {\bibinfo {volume} {105}},\ \bibinfo {pages}
  {015702} (\bibinfo {year} {2010})}\BibitemShut {NoStop}%
\bibitem [{\citenamefont {Ferretti}\ \emph {et~al.}(2010)\citenamefont
  {Ferretti}, \citenamefont {Andreani}, \citenamefont {T\"{u}reci},\ and\
  \citenamefont {Gerace}}]{Ferretti2010}%
  \BibitemOpen
  \bibfield  {author} {\bibinfo {author} {\bibfnamefont {S.}~\bibnamefont
  {Ferretti}}, \bibinfo {author} {\bibfnamefont {L.~C.}\ \bibnamefont
  {Andreani}}, \bibinfo {author} {\bibfnamefont {H.~E.}\ \bibnamefont
  {T\"{u}reci}}, \ and\ \bibinfo {author} {\bibfnamefont {D.}~\bibnamefont
  {Gerace}},\ }\href {\doibase 10.1103/PhysRevA.82.013841} {\bibfield
  {journal} {\bibinfo  {journal} {Phys. Rev. A}\ }\textbf {\bibinfo {volume}
  {82}},\ \bibinfo {pages} {013841} (\bibinfo {year} {2010})}\BibitemShut
  {NoStop}%
\bibitem [{\citenamefont {Lee}\ \emph {et~al.}(2011)\citenamefont {Lee},
  \citenamefont {H\"{a}ffner},\ and\ \citenamefont {Cross}}]{Lee2011a}%
  \BibitemOpen
  \bibfield  {author} {\bibinfo {author} {\bibfnamefont {T.~E.}\ \bibnamefont
  {Lee}}, \bibinfo {author} {\bibfnamefont {H.}~\bibnamefont {H\"{a}ffner}}, \
  and\ \bibinfo {author} {\bibfnamefont {M.~C.}\ \bibnamefont {Cross}},\ }\href
  {\doibase 10.1103/PhysRevA.84.031402} {\bibfield  {journal} {\bibinfo
  {journal} {Phys. Rev. A}\ }\textbf {\bibinfo {volume} {84}},\ \bibinfo
  {pages} {031402} (\bibinfo {year} {2011})}\BibitemShut {NoStop}%
\bibitem [{\citenamefont {Marcos}\ \emph {et~al.}(2012)\citenamefont {Marcos},
  \citenamefont {Tomadin}, \citenamefont {Diehl},\ and\ \citenamefont
  {Rabl.}}]{Marcos2012}%
  \BibitemOpen
  \bibfield  {author} {\bibinfo {author} {\bibfnamefont {D.}~\bibnamefont
  {Marcos}}, \bibinfo {author} {\bibfnamefont {A.}~\bibnamefont {Tomadin}},
  \bibinfo {author} {\bibfnamefont {S.}~\bibnamefont {Diehl}}, \ and\ \bibinfo
  {author} {\bibfnamefont {P.}~\bibnamefont {Rabl.}},\ }\href
  {http://iopscience.iop.org/1367-2630/14/5/055005} {\bibfield  {journal}
  {\bibinfo  {journal} {New Journal of Physics}\ }\textbf {\bibinfo {volume}
  {14}},\ \bibinfo {pages} {055005} (\bibinfo {year} {2012})}\BibitemShut
  {NoStop}%
\bibitem [{\citenamefont {Murch}\ \emph {et~al.}(2012)\citenamefont {Murch},
  \citenamefont {Vool}, \citenamefont {Zhou}, \citenamefont {Weber},
  \citenamefont {Girvin},\ and\ \citenamefont {Siddiqi}}]{Murch2012}%
  \BibitemOpen
  \bibfield  {author} {\bibinfo {author} {\bibfnamefont {K.~W.}\ \bibnamefont
  {Murch}}, \bibinfo {author} {\bibfnamefont {U.}~\bibnamefont {Vool}},
  \bibinfo {author} {\bibfnamefont {D.}~\bibnamefont {Zhou}}, \bibinfo {author}
  {\bibfnamefont {S.~J.}\ \bibnamefont {Weber}}, \bibinfo {author}
  {\bibfnamefont {S.~M.}\ \bibnamefont {Girvin}}, \ and\ \bibinfo {author}
  {\bibfnamefont {I.}~\bibnamefont {Siddiqi}},\ }\href
  {http://link.aps.org/doi/10.1103/PhysRevLett.109.183602} {\bibfield
  {journal} {\bibinfo  {journal} {Phys. Rev. Lett.}\ }\textbf {\bibinfo
  {volume} {109}},\ \bibinfo {pages} {183602} (\bibinfo {year}
  {2012})}\BibitemShut {NoStop}%
\bibitem [{\citenamefont {Lee}\ \emph {et~al.}(2012)\citenamefont {Lee},
  \citenamefont {H\"{a}ffner},\ and\ \citenamefont {Cross}}]{Lee2012}%
  \BibitemOpen
  \bibfield  {author} {\bibinfo {author} {\bibfnamefont {T.}~\bibnamefont
  {Lee}}, \bibinfo {author} {\bibfnamefont {H.}~\bibnamefont {H\"{a}ffner}}, \
  and\ \bibinfo {author} {\bibfnamefont {M.}~\bibnamefont {Cross}},\ }\href
  {\doibase 10.1103/PhysRevLett.108.023602} {\bibfield  {journal} {\bibinfo
  {journal} {Phys. Rev. Lett.}\ }\textbf {\bibinfo {volume} {108}},\ \bibinfo
  {pages} {023602} (\bibinfo {year} {2012})}\BibitemShut {NoStop}%
\bibitem [{\citenamefont {Grujic}\ \emph {et~al.}(2012)\citenamefont {Grujic},
  \citenamefont {Clark}, \citenamefont {Jaksch},\ and\ \citenamefont
  {Angelakis}}]{Grujic2012}%
  \BibitemOpen
  \bibfield  {author} {\bibinfo {author} {\bibfnamefont {T.}~\bibnamefont
  {Grujic}}, \bibinfo {author} {\bibfnamefont {S.~R.}\ \bibnamefont {Clark}},
  \bibinfo {author} {\bibfnamefont {D.}~\bibnamefont {Jaksch}}, \ and\ \bibinfo
  {author} {\bibfnamefont {D.~G.}\ \bibnamefont {Angelakis}},\ }\href {\doibase
  10.1088/1367-2630/14/10/103025} {\bibfield  {journal} {\bibinfo  {journal}
  {New J. Phys.}\ }\textbf {\bibinfo {volume} {14}},\ \bibinfo {pages} {103025}
  (\bibinfo {year} {2012})}\BibitemShut {NoStop}%
\bibitem [{\citenamefont {{Dalla Torre}}\ \emph {et~al.}(2012)\citenamefont
  {{Dalla Torre}}, \citenamefont {Demler}, \citenamefont {Giamarchi},\ and\
  \citenamefont {Altman}}]{Torre2011}%
  \BibitemOpen
  \bibfield  {author} {\bibinfo {author} {\bibfnamefont {E.~G.}\ \bibnamefont
  {{Dalla Torre}}}, \bibinfo {author} {\bibfnamefont {E.}~\bibnamefont
  {Demler}}, \bibinfo {author} {\bibfnamefont {T.}~\bibnamefont {Giamarchi}}, \
  and\ \bibinfo {author} {\bibfnamefont {E.}~\bibnamefont {Altman}},\ }\href
  {\doibase 10.1103/PhysRevB.85.184302} {\bibfield  {journal} {\bibinfo
  {journal} {Phys. Rev. B}\ }\textbf {\bibinfo {volume} {85}},\ \bibinfo
  {pages} {184302} (\bibinfo {year} {2012})}\BibitemShut {NoStop}%
\bibitem [{\citenamefont {{Dalla Torre}}\ \emph {et~al.}(2013)\citenamefont
  {{Dalla Torre}}, \citenamefont {Diehl}, \citenamefont {Lukin}, \citenamefont
  {Sachdev},\ and\ \citenamefont {Strack}}]{Torre2012}%
  \BibitemOpen
  \bibfield  {author} {\bibinfo {author} {\bibfnamefont {E.~G.}\ \bibnamefont
  {{Dalla Torre}}}, \bibinfo {author} {\bibfnamefont {S.}~\bibnamefont
  {Diehl}}, \bibinfo {author} {\bibfnamefont {M.~D.}\ \bibnamefont {Lukin}},
  \bibinfo {author} {\bibfnamefont {S.}~\bibnamefont {Sachdev}}, \ and\
  \bibinfo {author} {\bibfnamefont {P.}~\bibnamefont {Strack}},\ }\href
  {\doibase 10.1103/PhysRevA.87.023831} {\bibfield  {journal} {\bibinfo
  {journal} {Phys. Rev. A}\ }\textbf {\bibinfo {volume} {87}},\ \bibinfo
  {pages} {023831} (\bibinfo {year} {2013})}\BibitemShut {NoStop}%
\bibitem [{\citenamefont {{Le Boit\'{e}}}\ \emph {et~al.}(2013)\citenamefont
  {{Le Boit\'{e}}}, \citenamefont {Orso},\ and\ \citenamefont
  {Ciuti}}]{LeBoite2013a}%
  \BibitemOpen
  \bibfield  {author} {\bibinfo {author} {\bibfnamefont {A.}~\bibnamefont {{Le
  Boit\'{e}}}}, \bibinfo {author} {\bibfnamefont {G.}~\bibnamefont {Orso}}, \
  and\ \bibinfo {author} {\bibfnamefont {C.}~\bibnamefont {Ciuti}},\ }\href
  {http://link.aps.org/doi/10.1103/PhysRevLett.110.233601} {\bibfield
  {journal} {\bibinfo  {journal} {Phys. Rev. Lett.}\ }\textbf {\bibinfo
  {volume} {110}},\ \bibinfo {pages} {233601} (\bibinfo {year}
  {2013})}\BibitemShut {NoStop}%
\bibitem [{\citenamefont {Jin}\ \emph {et~al.}(2013)\citenamefont {Jin},
  \citenamefont {Rossini}, \citenamefont {Fazio}, \citenamefont {Leib},\ and\
  \citenamefont {Hartmann}}]{Leib2013}%
  \BibitemOpen
  \bibfield  {author} {\bibinfo {author} {\bibfnamefont {J.}~\bibnamefont
  {Jin}}, \bibinfo {author} {\bibfnamefont {D.}~\bibnamefont {Rossini}},
  \bibinfo {author} {\bibfnamefont {R.}~\bibnamefont {Fazio}}, \bibinfo
  {author} {\bibfnamefont {M.}~\bibnamefont {Leib}}, \ and\ \bibinfo {author}
  {\bibfnamefont {M.~J.}\ \bibnamefont {Hartmann}},\ }\href
  {http://link.aps.org/doi/10.1103/PhysRevLett.110.163605} {\bibfield
  {journal} {\bibinfo  {journal} {Phys. Rev. Lett}\ }\textbf {\bibinfo {volume}
  {110}},\ \bibinfo {pages} {163605} (\bibinfo {year} {2013})}\BibitemShut
  {NoStop}%
\bibitem [{\citenamefont {Lee}\ \emph {et~al.}(2013)\citenamefont {Lee},
  \citenamefont {Gopalakrishnan},\ and\ \citenamefont {Lukin}}]{Lee2013}%
  \BibitemOpen
  \bibfield  {author} {\bibinfo {author} {\bibfnamefont {T.~E.}\ \bibnamefont
  {Lee}}, \bibinfo {author} {\bibfnamefont {S.}~\bibnamefont {Gopalakrishnan}},
  \ and\ \bibinfo {author} {\bibfnamefont {M.~D.}\ \bibnamefont {Lukin}},\
  }\href {http://link.aps.org/doi/10.1103/PhysRevLett.110.257204} {\bibfield
  {journal} {\bibinfo  {journal} {Phys. Rev. Lett.}\ }\textbf {\bibinfo
  {volume} {110}},\ \bibinfo {pages} {257204} (\bibinfo {year}
  {2013})}\BibitemShut {NoStop}%
\bibitem [{\citenamefont {Genway}\ \emph {et~al.}(2013)\citenamefont {Genway},
  \citenamefont {Li}, \citenamefont {Ates}, \citenamefont {Lanyon},\ and\
  \citenamefont {Lesanovsky}}]{Genway2013}%
  \BibitemOpen
  \bibfield  {author} {\bibinfo {author} {\bibfnamefont {S.}~\bibnamefont
  {Genway}}, \bibinfo {author} {\bibfnamefont {W.}~\bibnamefont {Li}}, \bibinfo
  {author} {\bibfnamefont {C.}~\bibnamefont {Ates}}, \bibinfo {author}
  {\bibfnamefont {B.~P.}\ \bibnamefont {Lanyon}}, \ and\ \bibinfo {author}
  {\bibfnamefont {I.}~\bibnamefont {Lesanovsky}},\ }\href@noop {} {\  (\bibinfo
  {year} {2013})},\ \Eprint {http://arxiv.org/abs/1308.1424} {arXiv:1308.1424}
  \BibitemShut {NoStop}%
\bibitem [{\citenamefont {Hu}\ \emph {et~al.}(2013)\citenamefont {Hu},
  \citenamefont {{E. Lee}},\ and\ \citenamefont {{W. Clark}}}]{Hu}%
  \BibitemOpen
  \bibfield  {author} {\bibinfo {author} {\bibfnamefont {A.}~\bibnamefont
  {Hu}}, \bibinfo {author} {\bibfnamefont {T.}~\bibnamefont {{E. Lee}}}, \ and\
  \bibinfo {author} {\bibfnamefont {C.}~\bibnamefont {{W. Clark}}},\
  }\href@noop {} {\  (\bibinfo {year} {2013})},\ \Eprint
  {http://arxiv.org/abs/1305.2208} {arXiv:1305.2208} \BibitemShut {NoStop}%
\bibitem [{\citenamefont {Hartmann}\ \emph {et~al.}(2006)\citenamefont
  {Hartmann}, \citenamefont {Brand\~{a}o},\ and\ \citenamefont
  {Plenio}}]{Hartmann2006}%
  \BibitemOpen
  \bibfield  {author} {\bibinfo {author} {\bibfnamefont {M.~J.}\ \bibnamefont
  {Hartmann}}, \bibinfo {author} {\bibfnamefont {F.~G. S.~L.}\ \bibnamefont
  {Brand\~{a}o}}, \ and\ \bibinfo {author} {\bibfnamefont {M.~B.}\ \bibnamefont
  {Plenio}},\ }\href {\doibase 10.1038/nphys462} {\bibfield  {journal}
  {\bibinfo  {journal} {Nature Physics}\ }\textbf {\bibinfo {volume} {2}},\
  \bibinfo {pages} {849} (\bibinfo {year} {2006})}\BibitemShut {NoStop}%
\bibitem [{\citenamefont {Greentree}\ \emph {et~al.}(2006)\citenamefont
  {Greentree}, \citenamefont {Tahan}, \citenamefont {Cole},\ and\ \citenamefont
  {Hollenberg}}]{Greentree2006}%
  \BibitemOpen
  \bibfield  {author} {\bibinfo {author} {\bibfnamefont {A.~D.}\ \bibnamefont
  {Greentree}}, \bibinfo {author} {\bibfnamefont {C.}~\bibnamefont {Tahan}},
  \bibinfo {author} {\bibfnamefont {J.~H.}\ \bibnamefont {Cole}}, \ and\
  \bibinfo {author} {\bibfnamefont {L.~C.~L.}\ \bibnamefont {Hollenberg}},\
  }\href {\doibase 10.1038/nphys466} {\bibfield  {journal} {\bibinfo  {journal}
  {Nature Physics}\ }\textbf {\bibinfo {volume} {2}},\ \bibinfo {pages} {856}
  (\bibinfo {year} {2006})}\BibitemShut {NoStop}%
\bibitem [{\citenamefont {Angelakis}\ \emph {et~al.}(2007)\citenamefont
  {Angelakis}, \citenamefont {Santos},\ and\ \citenamefont
  {Bose}}]{Angelakis2007a}%
  \BibitemOpen
  \bibfield  {author} {\bibinfo {author} {\bibfnamefont {D.}~\bibnamefont
  {Angelakis}}, \bibinfo {author} {\bibfnamefont {M.}~\bibnamefont {Santos}}, \
  and\ \bibinfo {author} {\bibfnamefont {S.}~\bibnamefont {Bose}},\ }\href
  {\doibase 10.1103/PhysRevA.76.031805} {\bibfield  {journal} {\bibinfo
  {journal} {Phys. Rev. A}\ }\textbf {\bibinfo {volume} {76}},\ \bibinfo
  {pages} {031805} (\bibinfo {year} {2007})}\BibitemShut {NoStop}%
\bibitem [{\citenamefont {Hartmann}\ \emph {et~al.}(2008)\citenamefont
  {Hartmann}, \citenamefont {Brand\~{a}o},\ and\ \citenamefont
  {Plenio}}]{Hartmann2008}%
  \BibitemOpen
  \bibfield  {author} {\bibinfo {author} {\bibfnamefont {M.~J.}\ \bibnamefont
  {Hartmann}}, \bibinfo {author} {\bibfnamefont {F.~G. S.~L.}\ \bibnamefont
  {Brand\~{a}o}}, \ and\ \bibinfo {author} {\bibfnamefont {M.~B.}\ \bibnamefont
  {Plenio}},\ }\href {\doibase 10.1002/lpor.200810046} {\bibfield  {journal}
  {\bibinfo  {journal} {Laser and Photon. Rev.}\ }\textbf {\bibinfo {volume}
  {2}},\ \bibinfo {pages} {527} (\bibinfo {year} {2008})}\BibitemShut {NoStop}%
\bibitem [{\citenamefont {Schmidt}\ and\ \citenamefont
  {Koch}(2013)}]{Schmidt2013b}%
  \BibitemOpen
  \bibfield  {author} {\bibinfo {author} {\bibfnamefont {S.}~\bibnamefont
  {Schmidt}}\ and\ \bibinfo {author} {\bibfnamefont {J.}~\bibnamefont {Koch}},\
  }\href {\doibase 10.1002/andp.201200261} {\bibfield  {journal} {\bibinfo
  {journal} {Annalen der Physik}\ }\textbf {\bibinfo {volume} {525}},\ \bibinfo
  {pages} {395} (\bibinfo {year} {2013})}\BibitemShut {NoStop}%
\bibitem [{\citenamefont {Bardyn}\ and\ \citenamefont
  {Imamoglu}(2012)}]{Bardyn2012}%
  \BibitemOpen
  \bibfield  {author} {\bibinfo {author} {\bibfnamefont {C.-E.}\ \bibnamefont
  {Bardyn}}\ and\ \bibinfo {author} {\bibfnamefont {A.}~\bibnamefont
  {Imamoglu}},\ }\href {\doibase 10.1103/PhysRevLett.109.253606} {\bibfield
  {journal} {\bibinfo  {journal} {Phys. Rev. Lett.}\ }\textbf {\bibinfo
  {volume} {109}},\ \bibinfo {pages} {253606} (\bibinfo {year}
  {2012})}\BibitemShut {NoStop}%
\bibitem [{\citenamefont {Maziero}\ \emph {et~al.}(2010)\citenamefont
  {Maziero}, \citenamefont {Guzman}, \citenamefont {C\'{e}leri}, \citenamefont
  {Sarandy},\ and\ \citenamefont {Serra}}]{Maziero2010}%
  \BibitemOpen
  \bibfield  {author} {\bibinfo {author} {\bibfnamefont {J.}~\bibnamefont
  {Maziero}}, \bibinfo {author} {\bibfnamefont {H.~C.}\ \bibnamefont {Guzman}},
  \bibinfo {author} {\bibfnamefont {L.~C.}\ \bibnamefont {C\'{e}leri}},
  \bibinfo {author} {\bibfnamefont {M.~S.}\ \bibnamefont {Sarandy}}, \ and\
  \bibinfo {author} {\bibfnamefont {R.~M.}\ \bibnamefont {Serra}},\ }\href
  {\doibase 10.1103/PhysRevA.82.012106} {\bibfield  {journal} {\bibinfo
  {journal} {Phys. Rev. A}\ }\textbf {\bibinfo {volume} {82}},\ \bibinfo
  {pages} {012106} (\bibinfo {year} {2010})}\BibitemShut {NoStop}%
\bibitem [{\citenamefont {{\v{S}telmachovi\v{c} Peter}}\ and\ \citenamefont
  {{Bu\v{z}ek Vladim\'{\i}r}}(2004)}]{StelmachoviPeter2004}%
  \BibitemOpen
  \bibfield  {author} {\bibinfo {author} {\bibnamefont {{\v{S}telmachovi\v{c}
  Peter}}}\ and\ \bibinfo {author} {\bibnamefont {{Bu\v{z}ek Vladim\'{\i}r}}},\
  }\href {http://link.aps.org/doi/10.1103/PhysRevA.70.032313} {\bibfield
  {journal} {\bibinfo  {journal} {Phys. Rev. A}\ }\textbf {\bibinfo {volume}
  {70}},\ \bibinfo {pages} {032313} (\bibinfo {year} {2004})}\BibitemShut
  {NoStop}%
\bibitem [{\citenamefont {{Le Hur}}(2008)}]{Hur2008}%
  \BibitemOpen
  \bibfield  {author} {\bibinfo {author} {\bibfnamefont {K.}~\bibnamefont {{Le
  Hur}}},\ }\href {\doibase 10.1016/j.aop.2007.12.003} {\bibfield  {journal}
  {\bibinfo  {journal} {Ann. Phys}\ }\textbf {\bibinfo {volume} {323}},\
  \bibinfo {pages} {2208} (\bibinfo {year} {2008})}\BibitemShut {NoStop}%
\bibitem [{\citenamefont {Vidal}(2003)}]{Vidal2003}%
  \BibitemOpen
  \bibfield  {author} {\bibinfo {author} {\bibfnamefont {G.}~\bibnamefont
  {Vidal}},\ }\href {\doibase 10.1103/PhysRevLett.91.147902} {\bibfield
  {journal} {\bibinfo  {journal} {Phys. Rev. Lett.}\ }\textbf {\bibinfo
  {volume} {91}},\ \bibinfo {pages} {147902} (\bibinfo {year}
  {2003})}\BibitemShut {NoStop}%
\bibitem [{\citenamefont {Vidal}(2004)}]{Vidal2004}%
  \BibitemOpen
  \bibfield  {author} {\bibinfo {author} {\bibfnamefont {G.}~\bibnamefont
  {Vidal}},\ }\href {\doibase 10.1103/PhysRevLett.93.040502} {\bibfield
  {journal} {\bibinfo  {journal} {Phys. Rev. Lett.}\ }\textbf {\bibinfo
  {volume} {93}},\ \bibinfo {pages} {040502} (\bibinfo {year}
  {2004})}\BibitemShut {NoStop}%
\bibitem [{\citenamefont {White}(1992)}]{White:DMRG}%
  \BibitemOpen
  \bibfield  {author} {\bibinfo {author} {\bibfnamefont {S.~R.}\ \bibnamefont
  {White}},\ }\href@noop {} {\bibfield  {journal} {\bibinfo  {journal} {Phys.
  Rev. Lett.}\ }\textbf {\bibinfo {volume} {69}},\ \bibinfo {pages} {2863}
  (\bibinfo {year} {1992})}\BibitemShut {NoStop}%
\bibitem [{\citenamefont {White}(1993)}]{White:Algorithms}%
  \BibitemOpen
  \bibfield  {author} {\bibinfo {author} {\bibfnamefont {S.~R.}\ \bibnamefont
  {White}},\ }\href@noop {} {\bibfield  {journal} {\bibinfo  {journal} {Phys.
  Rev. B}\ }\textbf {\bibinfo {volume} {48}},\ \bibinfo {pages} {10345}
  (\bibinfo {year} {1993})}\BibitemShut {NoStop}%
\bibitem [{\citenamefont {Cazalilla}\ and\ \citenamefont
  {Marston}(2002)}]{Cazalilla2002}%
  \BibitemOpen
  \bibfield  {author} {\bibinfo {author} {\bibfnamefont {M.}~\bibnamefont
  {Cazalilla}}\ and\ \bibinfo {author} {\bibfnamefont {J.}~\bibnamefont
  {Marston}},\ }\href {\doibase 10.1103/PhysRevLett.88.256403} {\bibfield
  {journal} {\bibinfo  {journal} {Phys. Rev. Lett.}\ }\textbf {\bibinfo
  {volume} {88}},\ \bibinfo {pages} {256403} (\bibinfo {year}
  {2002})}\BibitemShut {NoStop}%
\bibitem [{\citenamefont {White}\ and\ \citenamefont
  {Feiguin}(2004)}]{White2004}%
  \BibitemOpen
  \bibfield  {author} {\bibinfo {author} {\bibfnamefont {S.~R.}\ \bibnamefont
  {White}}\ and\ \bibinfo {author} {\bibfnamefont {A.~E.}\ \bibnamefont
  {Feiguin}},\ }\href {\doibase 10.1103/PhysRevLett.93.076401} {\bibfield
  {journal} {\bibinfo  {journal} {Phys. Rev. Lett.}\ }\textbf {\bibinfo
  {volume} {93}},\ \bibinfo {pages} {076401} (\bibinfo {year}
  {2004})}\BibitemShut {NoStop}%
\bibitem [{\citenamefont {Daley}\ \emph {et~al.}(2004)\citenamefont {Daley},
  \citenamefont {Kollath}, \citenamefont {Schollw\"{o}ck},\ and\ \citenamefont
  {Vidal}}]{Daley2004}%
  \BibitemOpen
  \bibfield  {author} {\bibinfo {author} {\bibfnamefont {A.~J.}\ \bibnamefont
  {Daley}}, \bibinfo {author} {\bibfnamefont {C.}~\bibnamefont {Kollath}},
  \bibinfo {author} {\bibfnamefont {U.}~\bibnamefont {Schollw\"{o}ck}}, \ and\
  \bibinfo {author} {\bibfnamefont {G.}~\bibnamefont {Vidal}},\ }\href
  {\doibase 10.1088/1742-5468/2004/04/P04005} {\bibfield  {journal} {\bibinfo
  {journal} {J. Stat. Mech.}\ }\textbf {\bibinfo {volume} {2004}},\ \bibinfo
  {pages} {P04005} (\bibinfo {year} {2004})}\BibitemShut {NoStop}%
\bibitem [{\citenamefont {Micheli}\ \emph {et~al.}(2004)\citenamefont
  {Micheli}, \citenamefont {Daley}, \citenamefont {Jaksch},\ and\ \citenamefont
  {Zoller}}]{Micheli2004}%
  \BibitemOpen
  \bibfield  {author} {\bibinfo {author} {\bibfnamefont {A.}~\bibnamefont
  {Micheli}}, \bibinfo {author} {\bibfnamefont {A.}~\bibnamefont {Daley}},
  \bibinfo {author} {\bibfnamefont {D.}~\bibnamefont {Jaksch}}, \ and\ \bibinfo
  {author} {\bibfnamefont {P.}~\bibnamefont {Zoller}},\ }\href {\doibase
  10.1103/PhysRevLett.93.140408} {\bibfield  {journal} {\bibinfo  {journal}
  {Phys. Rev. Lett.}\ }\textbf {\bibinfo {volume} {93}},\ \bibinfo {pages}
  {140408} (\bibinfo {year} {2004})}\BibitemShut {NoStop}%
\bibitem [{\citenamefont {Clark}\ and\ \citenamefont
  {Jaksch}(2004)}]{Clark2004}%
  \BibitemOpen
  \bibfield  {author} {\bibinfo {author} {\bibfnamefont {S.}~\bibnamefont
  {Clark}}\ and\ \bibinfo {author} {\bibfnamefont {D.}~\bibnamefont {Jaksch}},\
  }\href {\doibase 10.1103/PhysRevA.70.043612} {\bibfield  {journal} {\bibinfo
  {journal} {Phys. Rev. A}\ }\textbf {\bibinfo {volume} {70}},\ \bibinfo
  {pages} {043612} (\bibinfo {year} {2004})}\BibitemShut {NoStop}%
\bibitem [{\citenamefont {Cai}\ and\ \citenamefont {Barthel}(2013)}]{Cai2013}%
  \BibitemOpen
  \bibfield  {author} {\bibinfo {author} {\bibfnamefont {Z.}~\bibnamefont
  {Cai}}\ and\ \bibinfo {author} {\bibfnamefont {T.}~\bibnamefont {Barthel}},\
  }\href {http://link.aps.org/doi/10.1103/PhysRevLett.111.150403} {\bibfield
  {journal} {\bibinfo  {journal} {Phys. Rev. Lett.}\ }\textbf {\bibinfo
  {volume} {111}},\ \bibinfo {pages} {150403} (\bibinfo {year}
  {2013})}\BibitemShut {NoStop}%
\bibitem [{\citenamefont {Schollw\"{o}ck}(2011)}]{Schollwock2011a}%
  \BibitemOpen
  \bibfield  {author} {\bibinfo {author} {\bibfnamefont {U.}~\bibnamefont
  {Schollw\"{o}ck}},\ }\href
  {http://www.sciencedirect.com/science/article/pii/S0003491610001752}
  {\bibfield  {journal} {\bibinfo  {journal} {Ann. Phys.}\ }\textbf {\bibinfo
  {volume} {326}},\ \bibinfo {pages} {96} (\bibinfo {year} {2011})}\BibitemShut
  {NoStop}%
\bibitem [{\citenamefont {Zwolak}\ and\ \citenamefont
  {Vidal}(2004)}]{Zwolak2004}%
  \BibitemOpen
  \bibfield  {author} {\bibinfo {author} {\bibfnamefont {M.}~\bibnamefont
  {Zwolak}}\ and\ \bibinfo {author} {\bibfnamefont {G.}~\bibnamefont {Vidal}},\
  }\href {\doibase 10.1103/PhysRevLett.93.207205} {\bibfield  {journal}
  {\bibinfo  {journal} {Phys. Rev. Lett.}\ }\textbf {\bibinfo {volume} {93}},\
  \bibinfo {pages} {207205} (\bibinfo {year} {2004})}\BibitemShut {NoStop}%
\bibitem [{\citenamefont {Or\'{u}s}\ and\ \citenamefont
  {Vidal}(2008)}]{Orus2008}%
  \BibitemOpen
  \bibfield  {author} {\bibinfo {author} {\bibfnamefont {R.}~\bibnamefont
  {Or\'{u}s}}\ and\ \bibinfo {author} {\bibfnamefont {G.}~\bibnamefont
  {Vidal}},\ }\href {\doibase 10.1103/PhysRevB.78.155117} {\bibfield  {journal}
  {\bibinfo  {journal} {Phys. Rev. B}\ }\textbf {\bibinfo {volume} {78}},\
  \bibinfo {pages} {155117} (\bibinfo {year} {2008})}\BibitemShut {NoStop}%
\bibitem [{\citenamefont {Prosen}\ and\ \citenamefont
  {\v{Z}nidari\v{c}}(2009)}]{Prosen2009}%
  \BibitemOpen
  \bibfield  {author} {\bibinfo {author} {\bibfnamefont {T.}~\bibnamefont
  {Prosen}}\ and\ \bibinfo {author} {\bibfnamefont {M.}~\bibnamefont
  {\v{Z}nidari\v{c}}},\ }\href {\doibase 10.1088/1742-5468/2009/02/P02035}
  {\bibfield  {journal} {\bibinfo  {journal} {J. Stat. Mech.}\ }\textbf
  {\bibinfo {volume} {2009}},\ \bibinfo {pages} {P02035} (\bibinfo {year}
  {2009})}\BibitemShut {NoStop}%
\bibitem [{\citenamefont {Fisher}\ \emph {et~al.}(1989)\citenamefont {Fisher},
  \citenamefont {Grinstein},\ and\ \citenamefont {Fisher}}]{Fisher1989}%
  \BibitemOpen
  \bibfield  {author} {\bibinfo {author} {\bibfnamefont {M.~P.~A.}\
  \bibnamefont {Fisher}}, \bibinfo {author} {\bibfnamefont {G.}~\bibnamefont
  {Grinstein}}, \ and\ \bibinfo {author} {\bibfnamefont {D.~S.}\ \bibnamefont
  {Fisher}},\ }\href {\doibase 10.1103/PhysRevB.40.546} {\bibfield  {journal}
  {\bibinfo  {journal} {Phys. Rev. B}\ }\textbf {\bibinfo {volume} {40}},\
  \bibinfo {pages} {546} (\bibinfo {year} {1989})}\BibitemShut {NoStop}%
\bibitem [{\citenamefont {Mattis}(2006)}]{mattis2006theory}%
  \BibitemOpen
  \bibfield  {author} {\bibinfo {author} {\bibfnamefont {D.~C.}\ \bibnamefont
  {Mattis}},\ }\href {http://books.google.co.uk/books?id=VkNBAQAAIAAJ} {\emph
  {\bibinfo {title} {{The Theory of Magnetism made simple}}}}\ (\bibinfo
  {publisher} {World Scientific},\ \bibinfo {address} {Singapore},\ \bibinfo
  {year} {2006})\BibitemShut {NoStop}%
\bibitem [{\citenamefont {Scully}\ and\ \citenamefont
  {Zubairy}(1997)}]{scully97}%
  \BibitemOpen
  \bibfield  {author} {\bibinfo {author} {\bibfnamefont {M.~O.}\ \bibnamefont
  {Scully}}\ and\ \bibinfo {author} {\bibfnamefont {M.~S.}\ \bibnamefont
  {Zubairy}},\ }\href@noop {} {\emph {\bibinfo {title} {{Quantum Optics}}}}\
  (\bibinfo  {publisher} {Cambridge University Press},\ \bibinfo {address}
  {Cambridge},\ \bibinfo {year} {1997})\BibitemShut {NoStop}%
\bibitem [{\citenamefont {Cresser}(1992)}]{Cresser1992a}%
  \BibitemOpen
  \bibfield  {author} {\bibinfo {author} {\bibfnamefont {J.}~\bibnamefont
  {Cresser}},\ }\href {\doibase 10.1080/09500349214552211} {\bibfield
  {journal} {\bibinfo  {journal} {J. Mod. Opt.}\ }\textbf {\bibinfo {volume}
  {39}},\ \bibinfo {pages} {2187} (\bibinfo {year} {1992})}\BibitemShut
  {NoStop}%
\bibitem [{\citenamefont {Joshi}\ \emph {et~al.}(2013)\citenamefont {Joshi},
  \citenamefont {Jonson}, \citenamefont {\"{O}hberg},\ and\ \citenamefont
  {Andersson}}]{Joshi2013}%
  \BibitemOpen
  \bibfield  {author} {\bibinfo {author} {\bibfnamefont {C.}~\bibnamefont
  {Joshi}}, \bibinfo {author} {\bibfnamefont {M.}~\bibnamefont {Jonson}},
  \bibinfo {author} {\bibfnamefont {P.}~\bibnamefont {\"{O}hberg}}, \ and\
  \bibinfo {author} {\bibfnamefont {E.}~\bibnamefont {Andersson}},\ }\href
  {http://link.aps.org/doi/10.1103/PhysRevA.87.062304} {\bibfield  {journal}
  {\bibinfo  {journal} {Phys. Rev. A}\ }\textbf {\bibinfo {volume} {87}},\
  \bibinfo {pages} {062304} (\bibinfo {year} {2013})}\BibitemShut {NoStop}%
\bibitem [{\citenamefont {Peres}(1996)}]{Peres1996}%
  \BibitemOpen
  \bibfield  {author} {\bibinfo {author} {\bibfnamefont {A.}~\bibnamefont
  {Peres}},\ }\href {\doibase 10.1103/PhysRevLett.77.1413} {\bibfield
  {journal} {\bibinfo  {journal} {Phys. Rev. Lett.}\ }\textbf {\bibinfo
  {volume} {77}},\ \bibinfo {pages} {1413} (\bibinfo {year}
  {1996})}\BibitemShut {NoStop}%
\bibitem [{\citenamefont {Horodecki}\ \emph {et~al.}(1996)\citenamefont
  {Horodecki}, \citenamefont {Horodecki},\ and\ \citenamefont
  {Horodecki}}]{Horodecki1996}%
  \BibitemOpen
  \bibfield  {author} {\bibinfo {author} {\bibfnamefont {M.}~\bibnamefont
  {Horodecki}}, \bibinfo {author} {\bibfnamefont {P.}~\bibnamefont
  {Horodecki}}, \ and\ \bibinfo {author} {\bibfnamefont {R.}~\bibnamefont
  {Horodecki}},\ }\href {\doibase 10.1016/S0375-9601(96)00706-2} {\bibfield
  {journal} {\bibinfo  {journal} {Physics Letters A}\ }\textbf {\bibinfo
  {volume} {223}},\ \bibinfo {pages} {1} (\bibinfo {year} {1996})}\BibitemShut
  {NoStop}%
\bibitem [{\citenamefont {Lupo}\ \emph {et~al.}(2005)\citenamefont {Lupo},
  \citenamefont {Man'ko}, \citenamefont {Marmo},\ and\ \citenamefont
  {Sudarshan}}]{Lupo2005}%
  \BibitemOpen
  \bibfield  {author} {\bibinfo {author} {\bibfnamefont {C.}~\bibnamefont
  {Lupo}}, \bibinfo {author} {\bibfnamefont {V.}~\bibnamefont {Man'ko}},
  \bibinfo {author} {\bibfnamefont {G.}~\bibnamefont {Marmo}}, \ and\ \bibinfo
  {author} {\bibfnamefont {E.}~\bibnamefont {Sudarshan}},\ }\href
  {http://iopscience.iop.org/0305-4470/38/48/009/} {\bibfield  {journal}
  {\bibinfo  {journal} {J. Phys. A: Math. Gen.}\ }\textbf {\bibinfo {volume}
  {38}},\ \bibinfo {pages} {009} (\bibinfo {year} {2005})}\BibitemShut
  {NoStop}%
\bibitem [{\citenamefont {Josza}\ and\ \citenamefont
  {Linden}(2003)}]{Josza2003}%
  \BibitemOpen
  \bibfield  {author} {\bibinfo {author} {\bibfnamefont {R.}~\bibnamefont
  {Josza}}\ and\ \bibinfo {author} {\bibfnamefont {N.}~\bibnamefont {Linden}},\
  }\href {http://rspa.royalsocietypublishing.org/content/459/2036/2011}
  {\bibfield  {journal} {\bibinfo  {journal} {Proc. R. Soc. A}\ }\textbf
  {\bibinfo {volume} {459}},\ \bibinfo {pages} {2011} (\bibinfo {year}
  {2003})}\BibitemShut {NoStop}%
\bibitem [{\citenamefont {Ollivier}\ and\ \citenamefont
  {Zurek}(2001)}]{Ollivier2001}%
  \BibitemOpen
  \bibfield  {author} {\bibinfo {author} {\bibfnamefont {H.}~\bibnamefont
  {Ollivier}}\ and\ \bibinfo {author} {\bibfnamefont {W.~H.}\ \bibnamefont
  {Zurek}},\ }\href {\doibase 10.1103/PhysRevLett.88.017901} {\bibfield
  {journal} {\bibinfo  {journal} {Phys. Rev. Lett.}\ }\textbf {\bibinfo
  {volume} {88}},\ \bibinfo {pages} {017901} (\bibinfo {year}
  {2001})}\BibitemShut {NoStop}%
\bibitem [{\citenamefont {Henderson}\ and\ \citenamefont
  {Vedral}(2001)}]{Henderson2001}%
  \BibitemOpen
  \bibfield  {author} {\bibinfo {author} {\bibfnamefont {L.}~\bibnamefont
  {Henderson}}\ and\ \bibinfo {author} {\bibfnamefont {V.}~\bibnamefont
  {Vedral}},\ }\href {\doibase 10.1088/0305-4470/34/35/315} {\bibfield
  {journal} {\bibinfo  {journal} {J. Phys. A.: Math. Gen}\ }\textbf {\bibinfo
  {volume} {34}},\ \bibinfo {pages} {6899} (\bibinfo {year}
  {2001})}\BibitemShut {NoStop}%
\bibitem [{\citenamefont {Modi}\ \emph {et~al.}(2012)\citenamefont {Modi},
  \citenamefont {Brodutch}, \citenamefont {Cable}, \citenamefont {Paterek},\
  and\ \citenamefont {Vedral}}]{Modi2012b}%
  \BibitemOpen
  \bibfield  {author} {\bibinfo {author} {\bibfnamefont {K.}~\bibnamefont
  {Modi}}, \bibinfo {author} {\bibfnamefont {A.}~\bibnamefont {Brodutch}},
  \bibinfo {author} {\bibfnamefont {H.}~\bibnamefont {Cable}}, \bibinfo
  {author} {\bibfnamefont {T.}~\bibnamefont {Paterek}}, \ and\ \bibinfo
  {author} {\bibfnamefont {V.}~\bibnamefont {Vedral}},\ }\href
  {http://link.aps.org/doi/10.1103/RevModPhys.84.1655} {\bibfield  {journal}
  {\bibinfo  {journal} {Rev. Mod. Phys.}\ }\textbf {\bibinfo {volume} {84}},\
  \bibinfo {pages} {1655} (\bibinfo {year} {2012})}\BibitemShut {NoStop}%
\bibitem [{\citenamefont {Knill}\ and\ \citenamefont
  {Laflamme}(1998)}]{Knill1998a}%
  \BibitemOpen
  \bibfield  {author} {\bibinfo {author} {\bibfnamefont {E.}~\bibnamefont
  {Knill}}\ and\ \bibinfo {author} {\bibfnamefont {R.}~\bibnamefont
  {Laflamme}},\ }\href {http://link.aps.org/doi/10.1103/PhysRevLett.81.5672}
  {\bibfield  {journal} {\bibinfo  {journal} {Phys. Rev. Lett.}\ }\textbf
  {\bibinfo {volume} {81}},\ \bibinfo {pages} {5672} (\bibinfo {year}
  {1998})}\BibitemShut {NoStop}%
\bibitem [{\citenamefont {Daki\'{c}}\ \emph {et~al.}(2012)\citenamefont
  {Daki\'{c}}, \citenamefont {Lipp}, \citenamefont {Ma}, \citenamefont
  {Ringbauer}, \citenamefont {Kropatschek}, \citenamefont {Barz}, \citenamefont
  {Paterek}, \citenamefont {Vedral}, \citenamefont {Zeilinger}, \citenamefont
  {Brukner},\ and\ \citenamefont {Walther}}]{Dakic2012a}%
  \BibitemOpen
  \bibfield  {author} {\bibinfo {author} {\bibfnamefont {B.}~\bibnamefont
  {Daki\'{c}}}, \bibinfo {author} {\bibfnamefont {Y.~O.}\ \bibnamefont {Lipp}},
  \bibinfo {author} {\bibfnamefont {X.}~\bibnamefont {Ma}}, \bibinfo {author}
  {\bibfnamefont {M.}~\bibnamefont {Ringbauer}}, \bibinfo {author}
  {\bibfnamefont {S.}~\bibnamefont {Kropatschek}}, \bibinfo {author}
  {\bibfnamefont {S.}~\bibnamefont {Barz}}, \bibinfo {author} {\bibfnamefont
  {T.}~\bibnamefont {Paterek}}, \bibinfo {author} {\bibfnamefont
  {V.}~\bibnamefont {Vedral}}, \bibinfo {author} {\bibfnamefont
  {A.}~\bibnamefont {Zeilinger}}, \bibinfo {author} {\bibfnamefont
  {C.}~\bibnamefont {Brukner}}, \ and\ \bibinfo {author} {\bibfnamefont
  {P.}~\bibnamefont {Walther}},\ }\href {http://dx.doi.org/10.1038/nphys2377}
  {\bibfield  {journal} {\bibinfo  {journal} {Nature Physics}\ }\textbf
  {\bibinfo {volume} {8}},\ \bibinfo {pages} {666} (\bibinfo {year}
  {2012})}\BibitemShut {NoStop}%
\bibitem [{\citenamefont {Daki\'{c}}\ \emph {et~al.}(2010)\citenamefont
  {Daki\'{c}}, \citenamefont {Vedral},\ and\ \citenamefont
  {Brukner}}]{Dakic2010}%
  \BibitemOpen
  \bibfield  {author} {\bibinfo {author} {\bibfnamefont {B.}~\bibnamefont
  {Daki\'{c}}}, \bibinfo {author} {\bibfnamefont {V.}~\bibnamefont {Vedral}}, \
  and\ \bibinfo {author} {\bibfnamefont {C.}~\bibnamefont {Brukner}},\ }\href
  {\doibase 10.1103/PhysRevLett.105.190502} {\bibfield  {journal} {\bibinfo
  {journal} {Phys. Rev. Lett.}\ }\textbf {\bibinfo {volume} {105}},\ \bibinfo
  {pages} {190502} (\bibinfo {year} {2010})}\BibitemShut {NoStop}%
\bibitem [{\citenamefont {Nissen}(2013)}]{Nissen2013a}%
  \BibitemOpen
  \bibfield  {author} {\bibinfo {author} {\bibfnamefont {F.~B.~F.}\
  \bibnamefont {Nissen}},\ }\emph {\bibinfo {title} {{Effects of Dissipation on
  Collective Behaviour in Circuit Quantum Electrodynamics}}},\ \href@noop {}
  {Ph.D. thesis},\ \bibinfo  {school} {University of Cambridge} (\bibinfo
  {year} {2013})\BibitemShut {NoStop}%
\bibitem [{\citenamefont {Hartmann}\ \emph {et~al.}(2009)\citenamefont
  {Hartmann}, \citenamefont {Prior}, \citenamefont {Clark},\ and\ \citenamefont
  {Plenio}}]{Hartmann2009}%
  \BibitemOpen
  \bibfield  {author} {\bibinfo {author} {\bibfnamefont {M.~J.}\ \bibnamefont
  {Hartmann}}, \bibinfo {author} {\bibfnamefont {J.}~\bibnamefont {Prior}},
  \bibinfo {author} {\bibfnamefont {S.~R.}\ \bibnamefont {Clark}}, \ and\
  \bibinfo {author} {\bibfnamefont {M.~B.}\ \bibnamefont {Plenio}},\ }\href
  {http://link.aps.org/doi/10.1103/PhysRevLett.102.057202} {\bibfield
  {journal} {\bibinfo  {journal} {Phys. Rev. Lett.}\ }\textbf {\bibinfo
  {volume} {102}},\ \bibinfo {pages} {057202} (\bibinfo {year}
  {2009})}\BibitemShut {NoStop}%
\bibitem [{\citenamefont {Bhaseen}\ \emph {et~al.}(2012)\citenamefont
  {Bhaseen}, \citenamefont {Mayoh}, \citenamefont {Simons},\ and\ \citenamefont
  {Keeling}}]{Bhaseen:Noneqdicke}%
  \BibitemOpen
  \bibfield  {author} {\bibinfo {author} {\bibfnamefont {M.~J.}\ \bibnamefont
  {Bhaseen}}, \bibinfo {author} {\bibfnamefont {J.}~\bibnamefont {Mayoh}},
  \bibinfo {author} {\bibfnamefont {B.~D.}\ \bibnamefont {Simons}}, \ and\
  \bibinfo {author} {\bibfnamefont {J.}~\bibnamefont {Keeling}},\ }\href
  {\doibase 10.1103/PhysRevA.85.013817} {\bibfield  {journal} {\bibinfo
  {journal} {Phys. Rev. A}\ }\textbf {\bibinfo {volume} {85}},\ \bibinfo
  {pages} {013817} (\bibinfo {year} {2012})}\BibitemShut {NoStop}%
\bibitem [{Note1()}]{Note1}%
  \BibitemOpen
  \bibinfo {note} {NB, the measure that we use, negativity, and that used in
  Refs.\cite {Osterloh2002,Osborne2002}, concurrence, are not identical. We
  have checked that plotting concurrence instead of negativity does not affect
  any of the conclusions discussed here.}\BibitemShut {Stop}%
\bibitem [{\citenamefont {Ma}\ \emph {et~al.}(2011)\citenamefont {Ma},
  \citenamefont {Wang}, \citenamefont {Sun},\ and\ \citenamefont
  {Nori}}]{Ma2011}%
  \BibitemOpen
  \bibfield  {author} {\bibinfo {author} {\bibfnamefont {J.}~\bibnamefont
  {Ma}}, \bibinfo {author} {\bibfnamefont {X.}~\bibnamefont {Wang}}, \bibinfo
  {author} {\bibfnamefont {C.~P.}\ \bibnamefont {Sun}}, \ and\ \bibinfo
  {author} {\bibfnamefont {F.}~\bibnamefont {Nori}},\ }\href
  {http://www.sciencedirect.com/science/article/pii/S0370157311002201}
  {\bibfield  {journal} {\bibinfo  {journal} {Physics Reports}\ }\textbf
  {\bibinfo {volume} {509}},\ \bibinfo {pages} {89} (\bibinfo {year}
  {2011})}\BibitemShut {NoStop}%
\bibitem [{\citenamefont {Yu}\ and\ \citenamefont {Eberly}(2009)}]{Yu2009}%
  \BibitemOpen
  \bibfield  {author} {\bibinfo {author} {\bibfnamefont {T.}~\bibnamefont
  {Yu}}\ and\ \bibinfo {author} {\bibfnamefont {J.~H.}\ \bibnamefont
  {Eberly}},\ }\href {\doibase 10.1126/science.1167343} {\bibfield  {journal}
  {\bibinfo  {journal} {Science}\ }\textbf {\bibinfo {volume} {323}},\ \bibinfo
  {pages} {598} (\bibinfo {year} {2009})}\BibitemShut {NoStop}%
\bibitem [{\citenamefont {Latorre}\ \emph {et~al.}(2004)\citenamefont
  {Latorre}, \citenamefont {Rico},\ and\ \citenamefont {Vidal}}]{Latorre2004a}%
  \BibitemOpen
  \bibfield  {author} {\bibinfo {author} {\bibfnamefont {J.~I.}\ \bibnamefont
  {Latorre}}, \bibinfo {author} {\bibfnamefont {E.}~\bibnamefont {Rico}}, \
  and\ \bibinfo {author} {\bibfnamefont {G.}~\bibnamefont {Vidal}},\ }\href
  {http://authors.library.caltech.edu/24913/} {\bibfield  {journal} {\bibinfo
  {journal} {Quantum Information and Computation}\ }\textbf {\bibinfo {volume}
  {4}},\ \bibinfo {pages} {48} (\bibinfo {year} {2004})}\BibitemShut {NoStop}%
\end{thebibliography}
%\bibliographystyle{apsrev4-1}

%

\end{document}